\documentclass[aoas]{imsart}

\RequirePackage{amsthm,amsmath,amsfonts,amssymb}
\RequirePackage[authoryear]{natbib}
\RequirePackage[colorlinks,citecolor=blue,urlcolor=blue]{hyperref}
\RequirePackage{graphicx}
\RequirePackage{times}
\usepackage{subcaption}
\startlocaldefs

\theoremstyle{remark}

\newcommand{\pkg}[1]{{\normalfont\fontseries{b}\selectfont #1}} 

\endlocaldefs

\begin{document}
	
	\begin{frontmatter}
		\title{Hierarchical Bayesian Modeling of Ocean Heat Content and its Uncertainty}
		\runtitle{Bayesian Ocean Heat Content}
		
		\begin{aug}
			\author[A]{\fnms{Samuel} \snm{Baugh}\ead[label=e1,mark]{samuelbaugh@ucla.edu}}
			\and
			\author[A,B]{\fnms{Karen} \snm{McKinnon}}
			
			\address[A]{Department of Statistics,
				UCLA,
				\printead{e1}}
			
			\address[B]{Institute for the Environment and Sustainability,
				UCLA}
		\end{aug}

		\begin{abstract}
			The accurate quantification of changes in the heat content of the world’s oceans is crucial for our understanding of the effects of increasing greenhouse gas concentrations. The Argo program, consisting of Lagrangian floats that measure vertical temperature profiles throughout the global ocean, has provided a wealth of data from which to estimate ocean heat content. However, creating a globally consistent statistical model for ocean heat content remains challenging due to the need for a globally valid covariance model that can capture complex nonstationarity. In this paper, we develop a hierarchical Bayesian Gaussian process model that uses kernel convolutions with cylindrical distances to allow for spatial non-stationarity in all model parameters while using a Vecchia process to remain computationally feasible for large spatial datasets. Our approach can produce valid credible intervals for globally integrated quantities that would not be possible using previous approaches. These advantages are demonstrated through the application of the model to Argo data, yielding credible intervals for the spatially varying trend in ocean heat content that accounts for both the uncertainty induced from interpolation and from estimating the mean field and other parameters. Through cross-validation, we show that our model out-performs an out-of-the-box approach as well as other simpler models. The code for performing this analysis is provided as the R package \pkg{BayesianOHC}.
		\end{abstract}
		
		
	\end{frontmatter}
	%
	%
	%

	\section{Introduction}\label{sec1}

	As 90\% of the energy imbalance caused by increased greenhouse gas concentrations is expected to be stored in the world's oceans \citep{trenberth2014earth}, measuring changes in ocean heat content is essential to the scientific understanding of climate change. However, obtaining accurate estimates of ocean heat content has historically been difficult due to a lack of data. To help address this problem, since 2002 the Argo program has been coordinating and maintaining a network of thousands of Lagrangian floats that passively move throughout the ocean measuring temperature and salinity \citep{Riser2016}. While Argo floats achieve dense sampling in the vertical dimension, quantifying ocean heat content remains challenging since the number of Argo observations is small in comparison to the vastness of the global ocean. As estimates of ocean heat content inform scientific knowledge about other climate processes, it is particularly important to not just obtain an estimate of the ocean heat content trend but to assess the uncertainty in its estimation. 
	
	Several techniques have been used in the climate science literature to quantify ocean heat content using Argo data. One group of techniques is referred to as simple-gridding \citep{meyssignac2019measuring} and includes approaches such as \citet{schuckmann2011well} and \citet{gouretski2018world}. These methods compute ocean heat content at each point of a grid over the ocean's surface by taking the average value of the observations within the corresponding grid-box. For grid-boxes with few observations, the heat content value is assumed to either have an anomaly value of zero or be equal to the global average value of all observations. Standard errors are obtained by computing the empirical standard deviation of the observations in each grid-box; for grid-boxes with few observations, the global standard deviation over all observations is used. A major drawback of this method is that it does not take advantage of the correlation structure in the data, which will lead to more variable estimates of the mean field and precludes a globally consistent estimate of the uncertainty in heat content. 
	
	Another group of approaches, including \citet{levitus2012world} and \citet{ishii2017accuracy}, estimate the heat content anomaly at a particular point using values observed at nearby locations. \citet{levitus2012world} estimate heat content with a weighted sum of observations within a fixed radius, where the weights decrease with the distance between the observation locations at a rate determined by a fixed correlation length scale. \cite{ishii2017accuracy} estimate unobserved values by minimizing a cost function that accounts for the influence of nearby observations by using an inverse-distance weighting also with a fixed correlation length scale. A major drawback of these techniques is that the correlation length scales are not informed by the data and must be chosen in advance of the interpolation. Standard errors are estimated in these techniques by computing empirical variances and covariances on the observations partitioned into grid-boxes. While this standard error calculation takes into account the fact that observations in nearby grid-boxes will be correlated, the standard errors do not take into account the spatial distribution of observations within each grid-box. Furthermore, like the simple gridding approach, this method does not produce a consistent covariance estimate across the global domain. 
	
	Other approaches use data from climate models to aid in the estimation of ocean heat content. The approach of \citet{cheng2016benefits} interpolates the data using covariances obtained from output of the Coupled Model Intercomparison Project Phase 5. Approaches such as \cite{balmaseda2013distinctive} interpolate the heat content field using climate model runs constrained by the observed values. Both of these methods suffer from the drawback that they are subject to the effect of climate model error on the covariance structure which can be difficult to assess and correct for \citep{palmer2017ocean}.
	
	Gaussian processes are a powerful technique for expressing a spatial field with a probability model. The key assumption of the Gaussian process is that every linear combination of values in a spatial field follows a multivariate normal distribution. Then the correlation between each pair of observations is determined by their locations as well as by parameters that can be estimated from the data. Under the Gaussian process assumption, the value at an unobserved location conditioned on a set of observed values has a normal distribution whose mean is the linear combination of the observations \citep{steinbook}. The uncertainty in estimating the value at the unobserved location can then be obtained through the standard deviation of the conditional normal distribution.
	
	The Gaussian process interpolation is similar to the weighted sum method used in \citet{levitus2012world} in that the estimates are given by a linear combination of the observations. However, the Gaussian process imposes a probability model on the spatial field, from which it can be shown mathematically that the conditional mean at an unobserved location is the best linear unbiased predictor \citep{steinbook}. Another advantage of assigning a probability model to the spatial field is that it allows for the correlation parameters to be determined through statistical inference, rather than chosen in an ad-hoc fashion. The uncertainty in the estimation of the parameters can also be assessed using statistical inference techniques.
	
	Recent work in the statistics literature by \cite{kuusela} uses Gaussian processes on the Argo data to assess variability in the temperature fields at fixed depths. One challenge of applying Gaussian processes is that they were originally developed for fields whose covariance parameters are constant over the domain. However, the statistical properties of the ocean temperature field are known to vary substantially between different regions of the global domain \citep{Chenge1601545}. \citet{kuusela} account for non-stationarity by assuming local stationarity within windows of a fixed size centered around each point of a grid partitioning the ocean. This allows the model to be fit independently at each grid-point yielding an estimate and uncertainty range for the value at the center of each window. However, as ocean heat content is a globally integrated quantity, assessing its uncertainty cannot be accomplished with this local method. Instead, a covariance model that is valid over the entire domain is required, which is the focus of this work.
	
	Rather than modeling temperature itself, we model ocean heat content, defined as the integral of the three-dimensional temperature field at a particular time \citep{heatcontentdef}. Thus, the problem is reduced to fitting a non-stationary model to the surface of the sphere. Even so, modeling non-stationary fields on a sphere is a major theoretical and computational challenge \citep{jeong2017spherical}. Kernel-convolution methods introduced by \cite{higdon} and further developed by \cite{Paciorek2006} and \cite{risser2016nonstationary} provide a flexible way to construct a Gaussian process with parameters that vary over the domain. In these methods, each location in the field is assigned a parameterized kernel function and the global model is obtained by computing convolutions between the kernels at any two points. When the kernel parameters vary smoothly, they describe the local behavior of the process at that point, yielding a clear interpretation of their meaning.
	
	While the convolutions have easily expressed forms when the domain is assumed to be Euclidean, they are more challenging to compute when the domain is spherical. Computing convolutions using numerical approximation of the integrals as in \cite{jeong2015class} and \cite{convsphere} is computationally intensive and even with pre-computation is infeasible on the thousands of observations in the Argo dataset. Various other non-stationary correlation functions have been developed for Gaussian processes on the sphere, such as deformations of the sphere introduced by \citet{sampson1992nonparametric}, spherical harmonics used by \cite{aniso}, and stochastic partial differential equations used by \cite{lindgren2011explicit}, however, the parameters of these covariance functions do not have the intuitive physical interpretation afforded by kernel convolution methods. Recently the \pkg{BayesNSGP} package (\cite{turek_bayesnsgp_2019}, \cite{risser_bayesian_2020}, and \cite{risser_nonstationary_2020}) offers the ability to use kernel convolutions in a Bayesian framework to model highly non-stationary processes in a computationally efficient manner. This package allows for non-stationary anisotropic modeling on the two-dimensional Euclidean domain and non-stationary isotropic modeling on the spherical domain.
	
	In this paper, we develop a kernel-convolution Gaussian process treating the surface of the ocean as a cylinder to produce a model that has spatially varying and interpretable parameters while remaining computationally feasible. We represent the spatially varying parameter fields themselves as Gaussian processes within a Bayesian hierarchical model structure. The Gaussian process prior for the parameters allows for the longitudinal correlation length scale to vary with respect to the changing radius of the Earth with latitude; this, in combination with the fact that there are no Argo observations close to the poles, allows for the use of the cylindrical distance rather than more computationally challenging spherical distance metric. Our methodology, including the covariance model, the Bayesian hierarchical framework, and the MCMC model-fitting procedure is described in detail in Section \ref{modelFramework}. Fitting this highly flexible Bayesian model on the number of points in the Argo dataset is still a computational challenge, and as such we use a Vecchia process as developed in \cite{katzfuss2021general} to improve the computation time. The accuracy of the Vecchia approximation for the kernel-convolution framework is demonstrated through comparison results on subsets of the data in Section \ref{vecchiaProcess}. The results of our model fit to the Argo data restricted to January is displayed in Section \ref{results}, demonstrating its ability to compute confidence and credible intervals for the parameter fields, ocean heat content, and the heat content trend over time. Section \ref{modelValidation} shows that our model out-performs an out-of-the-box method at predicting ocean heat content through cross-validation, and also shows the relative importance of the various aspects of the model. The functions used in performing this analysis are available as the R package \pkg{BayesianOHC}.

	\section{Data} 
	
	We first describe the data used for our statistical model. After deployment, an Argo float proceeds through four stages. First, the float adjusts its buoyancy to descend to $1{,}000$m of depth. In the second stage, the float drifts at $1{,}000$m of depth for an average of nine days, after which it descends further to $2{,}000$m. The float then ascends from $2{,}000$m to the surface, measuring temperature, salinity, and pressure at depth intervals averaging between $50$ and $100$ meters. Once the float surfaces, the measurements as well as the float's location and time at surfacing are transmitted via satellite \citep{Riser2016}.
	
	For computational simplicity, we restrict our analysis to profiles observed during January, although the computationally efficient Vecchia process described in Section \ref{vecchiaProcess} will allow for the model to be extended to all months in future work. As the focus of this work is a globally consistent spatial model, we do not model any temporal dependence in the data and implicitly take each data point as occurring at the mid-point of the month. We restrict our attention to profiles observed between $2007$ and $2016$ as this is the range of years included in the Argo data obtained from \cite{kuusela}, although an extension to additional years would be straightforward. For various reasons, some of the observations in the dataset were not recorded to the full depth of $2{,}000$m, and the percentage of floats with a maximum depth greater than $1{,}900$m ranges from $50\%$ in 2007 to $70\%$ in 2016. To avoid having to extrapolate the temperature field over depth, we exclude all floats with a maximum recorded depth shallower than $1{,}900$m. In total, the data under consideration consists of $42{,}776$ profiles, ranging from $2{,}581$ in 2007 to $7{,}543$ in 2016.

	\begin{figure}
		\centering
		
		\begin{subfigure}{\textwidth}
			\centering
			\includegraphics[width=.7\linewidth]{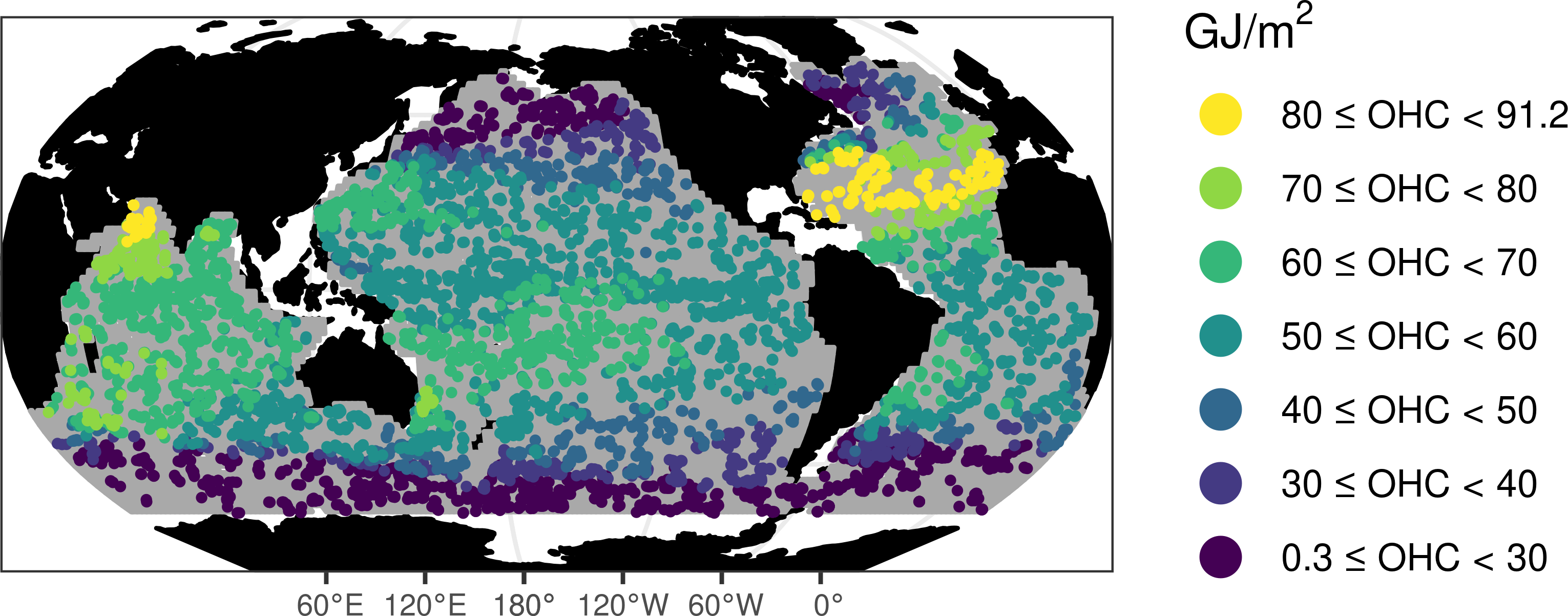}
			\caption{Vertical column ocean heat content values measured by Argo floats during January 2016. The mask used as the domain for interpolating ocean heat content is shaded in gray.} 	\label{ohc_field:raw_values}
		\end{subfigure}

		\vspace{.1in}
		
		\begin{subfigure}{\textwidth}
			\centering
			\includegraphics[width=.7\linewidth]{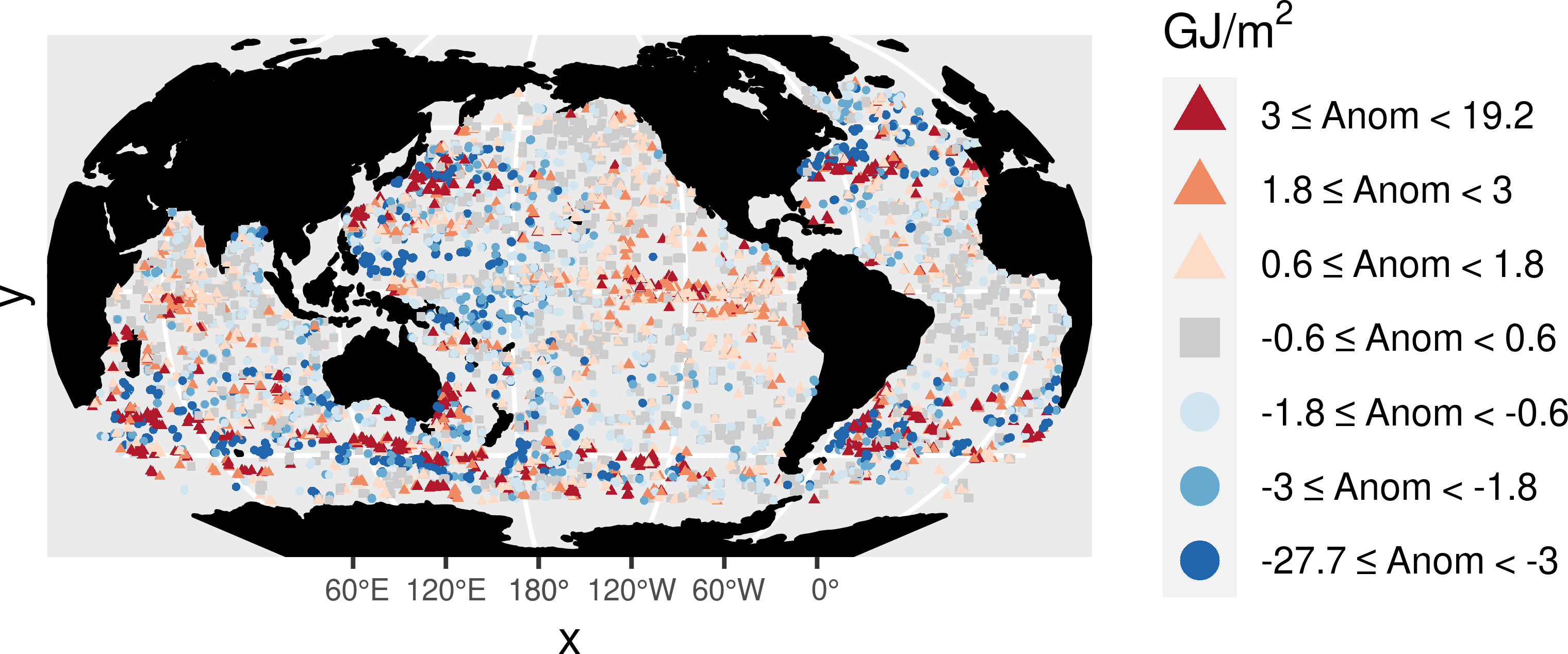}
			\caption{Observed heat content anomaly field, where anomalies are computed with respect to the spatially varying mean field estimated in section \ref{results}.} 	\label{ohc_field:anom_values}
		\end{subfigure}
		
		\caption{}\label{ohc_field}
	\end{figure}
	
	For notation, let $\textit{yr}\in\{2007,\dots,2016\}$ denote the year, $\textbf{x}=(x_{\textnormal{lat}},x_{\textnormal{lon}})\in\mathbb{S}^2$ denote spatial location in latitude-longitude coordinates, and $z\in\mathbb{R}^{+}$ denote depth from the surface in meters. Define the three-dimensional temperature field at location $\textbf{x}$, depth $z$, and year ${yr}$ at mid-January as $T_{\textit{yr}}(\textbf{x},z)$ measured in degrees Celsius. Since our ultimate interest is in the integrated heat content, we choose to model the two-dimensional spatial field of heat content values integrated over depth rather than the temperature and/or salinity fields at fixed depth levels.  The vertically integrated heat content field from 0-$2{,}000$m depth is defined as \begin{equation}
			H_{\textit{yr}}(\textbf{x})=\int_{0\textnormal{m}}^{2000\textnormal{m}} \rho c_pT_{\textit{yr}}(\textbf{x},z)\,dz \label{ohceq}
	\end{equation} where $\rho$ is seawater density in kg/m$^3$ and $c_p$ is the specific heat of seawater in J/(kg$^\circ$ C) \citep{heatcontentdef}. The units of $H_{\textit{yr}}(\textbf{x})$ are J/m$^2$. For observed temperature profiles the integral in Equation \ref{ohceq} is computed numerically by summing the values of the observations linearly interpolated to each meter of the domain [0m,2000m]. The heat content values as observed in January 2016 are shown in Figure \ref{ohc_field:raw_values}. Heat content anomalies computed with respect to the spatially varying mean field estimated in Section \ref{results} are shown in Figure \ref{ohc_field:anom_values}.

	The main quantity of interest in this work is the total ocean heat content or OHC, which is defined for a given year as the two-dimensional integral of the heat content field: \[\textnormal{OHC}_\textit{yr}=\int_{\textbf{x}\in S_{\textnormal{mask}}}H_{\textit{yr}}(\textbf{x})\,d\textbf{x}\]  where $S_{\textnormal{mask}}\subset\mathbb{S}^2$ is the ocean's surface masked following \citet{ROEMMICH200981}; this mask excludes regions of the ocean that are hard or impossible to sample with Argo floats, including regions of the ocean where the depth is less than $2{,}000$m, seasonally ice-covered regions, and marginal seas such as the Gulf of Mexico and the Mediterranean. The mask can be seen as the gray background of Figure \ref{ohc_field:raw_values}. The primary goal of this work is to estimate the trend in $\textnormal{OHC}_\textit{yr}$, and quantify the uncertainty in those estimates, using the observed Argo profiles.  Let $H_{\textnormal{obs};\textit{yr}}(\textbf{x})$ denote the discrete integral of the temperature profile observed at location $\textbf{x}$ and year $\textit{yr}$. In our modeling framework, the underlying heat content field $H_{\textit{yr}}$ differs from the observed heat content field $H_{\textnormal{obs};\textit{yr}}$ in that the latter contains small-scale variability that will be modeled as a nugget effect; see section \ref{modelFramework:covmod} for further details. In the following we denote the vector of observed values for year $yr\in\{2007,\dots,2016\}$ as $\textbf{H}_{\textnormal{obs};yr}$ and the concatenated vector of observed values as $\textbf{H}_{\textnormal{obs}}\equiv[\textbf{H}_{\textnormal{obs};2007},\dots,\textbf{H}_{\textnormal{obs};2016}]$.
	
	\section{Model Framework}\label{modelFramework} 
	
	We use a hierarchical Bayesian framework for our model of ocean heat content, in which both the spatial model parameters and the ocean heat content field are modeled as Gaussian processes. This approach yields several advantages. First, it provides a framework for propagating and quantifying uncertainty from both the inferred parameter fields and the spatial prediction of ocean heat content, thereby allowing us to determine the relative importance of each type of uncertainty. Second, the parameter fields are allowed to vary smoothly in space according to a Gaussian process with prescribed hyperparameters, without the need to choose a functional form. This flexibility is critical for modeling ocean heat content due to the complex structure of the anomaly field as can be seen in Figure \ref{ohc_field:anom_values}; this will be discussed quantitatively in Section \ref{modelValidation}.
	
	In this section, we will describe the non-stationary covariance model used for the heat content field, the hierarchical Bayesian framework including Gaussian process priors on the parameter fields, and the estimation of the posterior distribution of OHC using MCMC sampling with Metropolis-Hastings-within-Gibbs steps.
	
	\subsection{Covariance Model for the Heat Content Field}\label{modelFramework:covmod} 
	
	The heat content field is modeled as a Gaussian process with spatially varying covariance parameters in order to capture non-stationarity in the data. In the following, we denote the mean field as $\mu$, the variance field as $\phi$, the nugget variance field as $\sigma^2$, and the latitudinal and longitudinal correlation length scale fields as $\theta_{\textnormal{lat}}$ and $\theta_{\textnormal{lon}}$ respectively. For notation these symbols will refer to the fields as functions of location, so for example the value of the latitudinal correlation length scale at location \textbf{x} would be $\theta_{\textnormal{lat}}(\textbf{x})$. The field $\mu$ is defined for any \textbf{x} and year \textit{yr} as $\mu(\textbf{x})=\mu_{2007}(\textbf{x})+\beta(\textbf{x})(\textit{yr}-2007)$ where $\mu_{2007}$ is a spatially varying mean field centered on $2007$ and  $\beta$ is a spatially varying trend field. We will denote the collection of parameter fields as $\boldsymbol{\rho}\equiv\{\phi,\theta_{\textnormal{lat}},\theta_{\textnormal{lon}},\sigma^2,\mu_{2007},\beta\}$. The observed and underlying heat content fields are distributed as \[H_{\textnormal{obs};\textit{yr}}(\textbf{x})\overset{\textnormal{i.i.d.}}{\sim} N(H_{\textit{yr}}(\textbf{x}),\sigma^2(\textbf{x}))\] and \begin{equation}
			H_{\textit{yr}}(\textbf{x})\sim GP(\mu(\textbf{x}),k(\textbf{x},\textbf{y};
			\boldsymbol{\rho}))
			\label{fullHmodel}
	\end{equation} respectively where $GP$ is a Gaussian process such that $k(\textbf{x},\textbf{y};\boldsymbol{\rho})\equiv \textnormal{cov}[H_{yr}(\textbf{x}),H_{yr}(\textbf{y})]$ for any pair of locations $\textbf{x}$ and $\textbf{y}$. We make explicit the dependency of $k$ on the entire set of parameters $\boldsymbol{\rho}$ for notational simplicity, although the value of $k$ will depend only on the values of $\theta_{\textnormal{lat}}$, $\theta_{\textnormal{lon}}$, and $\phi$ for the two relevant locations.

	The nugget variance $\sigma^2$ is generally interpreted as measurement error, however as measurement errors in Argo temperature readings are known to be small \citep{Riser2016} it will be used here to represent fine-scaled variation in the underlying field. This can be induced, for example, by eddies, which have coherent structure but are generally smaller in spatial and temporal scale than the observation resolution of the Argo profiles \citep{colling2001ocean}. This form of variation is excluded from the kriging-based estimates of ocean heat content anomaly in Equation \ref{fullHmodel}, but is represented in the uncertainty estimates. In the model fit in Section \ref{results}, we sample the inverse signal-to-noise ratio $\sigma^2(\textbf{x})/\phi(\textbf{x})$ as this field appears to have better sampling properties, however for simplicity we describe here the analogous case where $\sigma^2$ is sampled directly.
	
	Kernel convolution methods as introduced by \citet{higdon} are a straightforward and intuitive way for integrating a spatially varying parameter field into a valid covariance model. The idea behind these methods is to assign a kernel to each point in the domain that gives the local covariance properties at that location. Then a valid global Gaussian process can be obtained by computing the convolution of the kernels between every two points. The correlation between any two points will reflect the local behavior at both of their locations.  \citet{Paciorek2006} demonstrated the use of kernel convolution methods and find closed-form expressions for the covariance functions when the underlying domain is Euclidean. For modeling the ocean heat content field it would make the most sense to measure distances between observations using the great-circle distance, which is the shortest distance between two points over the sphere's surface.  Unfortunately, the kernel convolutions do not have closed-form expressions when the great-circle distance is used, and the computational task of performing numerical integration of the convolution terms as done in \cite{convsphere} is not computationally feasible for the number of observations in the Argo data.  

		In principle, the spherical domain could be represented as a subset of three-dimensional Euclidean space and the chord-length distance, which is the shortest line between two points and is allowed to cross through the interior of the sphere, could be used. However, in $\mathbb{R}^3$ the traditional method of representing anisotropy through scaling and rotating the coordinate space  would not yield parameters that could be interpreted in terms of quantities on the surface. This is a limitation when modeling the ocean heat content field which, as we can see in Figure \ref{ohc_field}, exhibits significant anisotropy in the latitudinal and longitudinal dimensions. To examine the significance of this limitation we evaluate the ability of isotropic models to predict ocean heat content in Section \ref{modelValidation}.
	
	\begin{figure}
		\centering
		\includegraphics[scale=0.2]{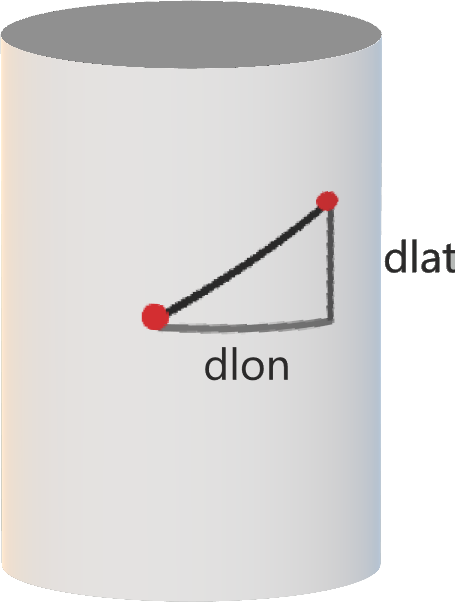}
		\caption{Visualization of the cylindrical distance metric. Here, \textit{dlat} is the Euclidean distance between the latitude coordinates and \textit{dlon} is the great circle distance between the longitude coordinates.}
		\label{cylinder}
	\end{figure}
	
	Our solution to the problem of modeling anisotropy on the sphere is to represent the Earth's surface in latitude-longitude coordinates as a cylinder, or $\mathbb{CL}=\mathbb{R}\times\mathbb{S}^{1}$, and use a cylindrical distance function for the covariance model that allows for the estimation of correlation length scales in the latitudinal and longitudinal dimensions. As Argo floats do not travel close to the poles, the approximation of the Earth's surface as a cylinder does not present any modeling problems. In particular, the geometric variation in the radius of the sphere with latitude can be accounted for in the estimation of the spatially varying longitudinal scale parameter. Let $d_{\textnormal{euc}}(x_{\textnormal{lat}},y_{\textnormal{lat}})$ denote the Euclidean distance between two points $x_{\textnormal{lat}},y_{\textnormal{lat}}\in\mathbb{R}$ and let $d_{\textnormal{gc}}(x_{\textnormal{lon}},y_{\textnormal{lon}})$ denote the great circle distance between two points $x_{\textnormal{lon}},y_{\textnormal{lon}}\in\mathbb{S}^1$. Then the shortest path between two locations on the cylinder is given by the Pythagorean theorem as visualized in Figure \ref{cylinder}. Latitudinal and longitudinal correlation length scale parameters can be incorporated in a straightforward way, yielding the following distance function, where $\textbf{x}=[x_{\textnormal{lat}},x_{\textnormal{lon}}]$ and  $\textbf{u}=[u_{\textnormal{lat}},u_{\textnormal{lon}}]$ are arbitrary locations: \[d_{\textnormal{cyl}}(\textbf{x},\textbf{u};\theta_{\textnormal{lat}},\theta_{\textnormal{lon}})=\sqrt{\frac{d_{\textnormal{euc}}(x_{\textnormal{lat}},{u}_{\textnormal{lat}})^2}{\theta_{\textnormal{lat}}(\textbf{x})}+\frac{d_{\textnormal{gc}}(x_{\textnormal{lon}},u_{\textnormal{lon}})^2}{\theta_{\textnormal{lon}}(\textbf{x})}}.\]
	Using this distance function in the kernel convolution framework with Gaussian kernels yields the following covariance function for arbitrary locations $\textbf{x}$ and $\textbf{y}$: \begin{equation}
			k(\textbf{x},\textbf{y};\boldsymbol{\rho})=\phi(\textbf{x})\phi(\textbf{y}) \int_{\textbf{u}\in\mathbb{CL}}\exp\left(-d_{\textnormal{cyl}}(\textbf{x},\textbf{u};\theta_{\textnormal{lat}},\theta_{\textnormal{lon}})^2-d_{\textnormal{cyl}}(\textbf{y},\textbf{u};\theta_{\textnormal{lat}},\theta_{\textnormal{lon}})^2\right)\,d\textbf{u}. \label{corrdef}
	\end{equation}
	By using Gaussian covariance kernels we can integrate over $\mathbb{CL}$ separately for the latitudinal and longitudinal dimensions. The integral over the Euclidean (latitudinal) dimension can then be computed easily through the closed-form expressions of \citet{Paciorek2006}. The convolutions in the cylindrical dimension can be computed exactly using Gaussian error functions as derived in Section S3.1 of the supplementary material. Supplementary Section S3.2 presents a faster Gaussian approximation to the exact spherical convolutions along with a simulation study demonstrating that the approximation is highly accurate over the correlation length scales present in the ocean heat content data. As such the approximation to the exact convolutions is used in the following analysis, although both the approximation and exact methods are available in the \pkg{BayesianOHC} package as "cylindrical\_correlation\_gaussian" and "cylindrical\_correlation\_exact" respectively. 
	
	What results from Equation \ref{corrdef} is a covariance function that can be locally anisotropic and globally non-stationary, is fast to compute, and is mathematically valid on the domain. A major advantage of the cylindrical distance model's anisotropy structure is that the range parameters are directly interpretable as correlation length scales in the latitudinal and longitudinal dimensions. To preserve this interpretability the cylindrical distance model does not allow for rotation of the anisotropy structure.

	\subsection{Hierarchical Bayesian Framework}\label{modelFramework:hierBayes}
	
	Having prescribed the structure of our covariance function for ocean heat content, we now turn to our model for the covariance parameters, which are themselves modeled as Gaussian processes. Each parameter field has a stationary Gaussian process prior with mean, variance, and spatial range hyperparameters. Fitting this complex model is made computationally feasible through the Vecchia process described in Section \ref{vecchiaProcess}. 
	
	In the following, let $\Sigma_{\textnormal{obs},\textnormal{obs};\boldsymbol{\rho}}$ denote the covariance matrix between each pair of observations as computed by $k$. Because we will ultimately fit the model to only data in a single month, we treat each year as independent, and in this case $\Sigma_{\textnormal{obs},\textnormal{obs};\boldsymbol{\rho}}$ is a block diagonal matrix. Future work will incorporate temporal dependence structures in the covariance matrix. Let $\rho\in\boldsymbol{\rho}$ denote an arbitrary parameter field.  Then the prior for each $\rho\in\boldsymbol{\rho}$ is a stationary Gaussian process with link function $g_\rho$, specifically \[g_\rho^{-1}(\rho(\textbf{x}))\sim GP(\mu_\rho,\phi_\rho\exp(-d_{\textnormal{cyl}}(\textbf{x},\textbf{y})/\theta_\rho))\] where $\mu_\rho$, $\phi_\rho,$ and $\theta_\rho$ are respectively the mean, variance, and range for the parameter field $\rho$. The link function $g_\rho$ is exponential for the variance and correlation parameters in order to ensure positivity, and is the identity for the mean and trend parameters. To constrain the number of parameters involved in representing the spatial surfaces, we will represent each by independent normal basis values $\textbf{b}_\rho\sim N(0,I_k)$ on a set of knot locations denoted ${\kappa}$; let $n_{\kappa}$ refer to the number of knot points. Specifically, for each $\rho$ we define the vector $\textbf{b}_\rho$ to be such that for any location $\textbf{x}$ the following holds: \begin{equation}
			\rho(\textbf{x})=g_\rho(\mu_\rho+\phi_\rho \Sigma_{\textbf{x},\kappa;\theta_\rho}\Sigma^{-1/2}_{\kappa,\kappa;\theta_{\rho}}\textbf{b}_\rho) \label{krigrho}
	\end{equation} where $\Sigma_{\textbf{x},\kappa;\theta_\rho}$ is the covariance vector between $\textbf{x}$ and the knot locations $\kappa$ and $\Sigma_{\kappa,\kappa;\theta_\rho}$ is the covariance matrix between points in $\kappa$. Intuitively, $\textbf{b}_\rho$ is a de-meaned and de-correlated representation of $g^{-1}(\rho(\textbf{x}))$ evaluated at the knot locations, from which the field can be re-constructed through kriging. In the model fit in Section \ref{results}, we choose $\kappa$ to contain knot locations at every $16$ degrees in the longitudinal direction and every $8$ degrees in the latitudinal direction. Restricting the knot locations with the data mask expressed in Figure \ref{ohc_field:raw_values}, this results in $n_{\kappa}=253$ basis values for each parameter field. This knot selection is somewhat arbitrary, and more principled methods for selecting knot point locations have been explored, such as \citet{katzfuss2013bayesian} who incorporate the knot locations into the Bayesian hierarchy through a reversible-jump process, however such approaches are not considered here.
	
	\subsection{MCMC Sampling Procedure}\label{modelFramework:mcmcSamp}
	
	We will use Markov Chain Monte Carlo (MCMC) with Gibbs steps for each of the parameter fields to estimate the joint posterior distribution.  The marginal distributions for the correlation length and variance parameters must be approximated using Metropolis-Hastings steps, however $\mu_{2007}$ and $\beta$ can be sampled jointly from their Gibbs marginal distribution given by the formula \begin{equation}[\textbf{b}_{\mu_{2007}},\textbf{b}_{\beta}]|\textbf{H}_{\textnormal{obs}},\textbf{b}_{\boldsymbol{\rho}\setminus\{\mu_{2007},\beta\}}\sim N(\mu_{\mu;\textnormal{post}},\Sigma_{\mu;\textnormal{post}})\label{beta_post} 
	\end{equation}  where
	\[\Sigma_{\mu;\textnormal{post}}=(M^T\Sigma_{\textnormal{obs},\textnormal{obs};\boldsymbol{\rho}}^{-1}M+I_{2n_{\kappa}})^{-1},\] \[\mu_{\mu;\textnormal{post}}=\Sigma_{\mu;\textnormal{post}}\left(M^T\Sigma_{\textnormal{obs,obs};\boldsymbol{\rho}}^{-1}\textbf{H}_{\textnormal{obs}}\right),\]  and $M\in\mathbb{R}^{n_{\textnormal{obs}},2n_{\kappa}}$ is the design matrix for $\textbf{b}_{\mu_{2007}}$ and $\textbf{b}_{\beta}$. In the conditional distribution of Equation \ref{beta_post}, "$\setminus$" is the set minus operator and as such $\textbf{b}_{\boldsymbol{\rho}\setminus\{\mu_{2007},\beta\}}$ denotes the set of basis parameters for all of the parameter fields excluding $\mu_{2007}$ and $\beta$.
	
	Now we will outline the MCMC sampling procedure.  For each iteration $t>1$ after initialization we perform the following steps:
	
	\begin{enumerate}
		\item For a given $\rho\in\{\theta_{\textnormal{lat}},\theta_{\textnormal{lon}},\sigma^2,\phi\}$, propose basis point candidates through independent perturbations, specifically $\textbf{b}_\rho^{(t)}\sim N\left(\textbf{b}_\rho^{(t-1)},\left(\sigma^{(t)}_{\rho;\textnormal{prop}}\right)^2 I_{n_{\kappa}}\right)$ where $\sigma^{(t)}_{\rho;\textnormal{prop}}$ is the proposal standard deviation at the current iteration. The independent perturbations on $\textbf{b}_\rho$ can be viewed as perturbing $\log\rho^{(t)}$ by a spatially correlated field that shares the spatial range of $\log\rho^{(t)}$.
		\item Accept or reject the entire basis vector, which describes the full spatial field, using the Metropolis-Hastings acceptance ratio on  \[P(\textbf{b}^{(t)}_\rho|\textbf{H}_{\textnormal{obs}},\textbf{b}_{\boldsymbol{\rho}\setminus\{\rho\}})\propto P(\textbf{H}_{\textnormal{obs}}|\rho^{(t)}(\textbf{x}_{\textnormal{obs}}),\textbf{b}_{\boldsymbol{\rho}})P(\textbf{b}^{(t)}_\rho)\] where $\textbf{b}_{\boldsymbol{\rho}\setminus\{\rho\}}$ denotes the set of basis points corresponding to all of the fields except for $\rho$ at their most recent iteration and $\textbf{x}_{\textnormal{obs}}$ denotes the full set of observation locations. The values of $\rho$ at the observation locations can be evaluated through Equation $\ref{krigrho}$. Here, the data likelihood term on the right is multivariate Gaussian with covariance matrix $\Sigma_{\textnormal{obs,obs};\boldsymbol{\rho}}$ and the prior term $P(\textbf{b}^{(t)}_\rho)$ is the product of independent standard normal densities.
		\item Repeat the previous two steps for sampling $\textbf{b}_{\rho}^{(t)}$ for each remaining $\rho\in\{\theta_{\textnormal{lat}},\theta_{\textnormal{lon}},\sigma^2,\phi\}$.
		\item Sample $\textbf{b}_{\mu_{2007}}^{(t)}$ and $\textbf{b}_{\beta}^{(t)}$ jointly from their conditional distribution given in Equation \ref{beta_post}.
		\item Update the variances $\sigma_{\rho;\textnormal{prop}}^{(t)}$ using the adaptive MCMC algorithm for Metropolis-within-Gibbs sampling as described in \citet{rosen}, which updates the proposal variances to achieve a desired acceptance rate chosen to be $.44$.
	\end{enumerate}
	
	Through this approach, MCMC samples of coherent spatial fields for each model parameter can be obtained. In particular, we model the mean field parameters jointly with the covariance parameters, which will generally lead to improved estimates of both. Further, our resulting credible intervals will incorporate the uncertainty induced by estimating the mean and the spatially varying trend. Quantifying the trend is particularly valuable as for many purposes we are primarily concerned with the trend in ocean heat content anomaly over time rather than the values of the anomalies themselves.
	
	\subsection{Posterior Distributions}\label{modelFramework:posterior}
	
	For a fixed set of parameter values sampled from the posterior, the distribution of the globally-integrated ocean heat content, $\textnormal{OHC}_\textit{yr}$, is normal and can be denoted $\textnormal{OHC}_\textit{yr}|\boldsymbol{\rho},\textbf{x}_{\textit{yr};\textnormal{obs}}\sim N(\mu_{\textnormal{OHC}_\textit{yr}},\sigma^2_{\textnormal{OHC}_\textit{yr}})$ with values \[\mu_{\textnormal{OHC}_\textit{yr}}=\Sigma_{\textnormal{globe,obs};\boldsymbol{\rho}}\Sigma_{\textnormal{obs,obs};\boldsymbol{\rho}}\textbf{H}_{\textnormal{obs};\textit{yr}}\textnormal{  and}\]    \begin{equation}
			\sigma_{\textnormal{OHC}_{\textit{yr}}}^2=\Sigma_{\textnormal{globe,globe};\boldsymbol{\rho}}-\Sigma_{\textnormal{globe,obs};\boldsymbol{\rho}}\Sigma_{\textnormal{obs,obs};\boldsymbol{\rho}}^{-1}\Sigma_{\textnormal{obs,globe};\boldsymbol{\rho}} \label{ohc_post_sd}
	\end{equation} where  $\Sigma_{\textnormal{globe,obs};\boldsymbol{\rho}}$ denotes the covariance vector between the observations locations and the global integral and $\Sigma_{\textnormal{globe},\textnormal{globe};\boldsymbol{\rho}}$ denotes the variance of the global integral. When $\mu_{\textnormal{OHC}_\textit{yr}}$ and $\sigma_{\textnormal{OHC}_\textit{yr}}$ are computed from the sampled parameters $\boldsymbol{\rho}^{(t)}$ we will refer to them as $\mu_{\textnormal{OHC}_\textit{yr}}^{(t)}$ and $\sigma^{(t)}_{\textnormal{OHC}_\textit{yr}}$. The entries in these covariance matrices involve integrals over the global domain which must be computed numerically. Let $\textbf{x}_{\textnormal{grid}}$ denote the grid-points to be used for numeric integration; then we can write our approximation of ocean heat content as $\textnormal{OHC}_\textit{yr}=a_{\textnormal{globe}}^T\textbf{H}_{\textnormal{globe};\textit{yr}}$ where $\textbf{H}_{\textnormal{globe};\textit{yr}}$ are the unknown ocean heat content values at the grid-points and $a_{\textnormal{globe}}$ is the vector of grid-cell areas corresponding to $\textbf{x}_{\textnormal{globe}}$. Then the above formulas apply replacing $\Sigma_{\textnormal{globe},\textnormal{globe}}$ with $\Sigma_{\textnormal{grid,grid}}$. In the results reported in Section \ref{results}, OHC posterior distributions are calculated using a grid partitioning the domain with a resolution of $1^\circ\times 1^\circ$.

	\section{Vecchia Processes for Feasible Bayesian Estimation}\label{vecchiaProcess}
	
	The most computationally intensive step of the Metropolis-Hastings procedure is computing the log-likelihood term. The log-likelihood term can be written as \[\log P(\textbf{H}_{\textnormal{obs}}|\boldsymbol{\rho})\propto-\left(\textbf{H}_{\textnormal{obs}}-\mu(\textbf{x}_{\textnormal{obs}})\right)^T\Sigma_{\textnormal{obs},\textnormal{obs}}^{-1}\left(\textbf{H}_{\textnormal{obs}}-\mu(\textbf{x}_{\textnormal{obs}})\right)-\log\textnormal{det}(\Sigma_{\textnormal{obs},\textnormal{obs}}^{-1})\] where we omit the dependency of $\Sigma_{\textnormal{obs,obs}}$ on $\boldsymbol{\rho}$ for notational simplicity. The usual way to compute the inverse and log-determinant terms in the above equation is to first compute the Cholesky decomposition of the covariance matrix, resulting in the upper-triangular Cholesky factor $C$  which satisfies the property $C^TC=\Sigma_{\textnormal{obs},\textnormal{obs}}$. Then the likelihood evaluation becomes \begin{equation}\log P(\textbf{H}_{\textnormal{obs}}|\boldsymbol{\rho})\propto-\left(C^{-1}(\textbf{H}_{\textnormal{obs}}-\mu(\textbf{x}_{\textnormal{obs}})\right)^T(C^{-1}(\textbf{H}_{\textnormal{obs}}-\mu(\textbf{x}_{\textnormal{obs}})))-2\sum_{i=1}^n\log C_{ii}.\label{lik_chol}
	\end{equation}  Inverting an upper triangular matrix is computationally simple, so the primary burden is computing the Cholesky factorization. This step has a computational complexity of $O(n^3)$, meaning that the computation time increases cubically with the number of observations. For example, in a simple experiment on a standard laptop computer, computing the likelihood for the $2{,}581$ Argo observations in 2007 takes around $7$ seconds, while computing the likelihood for the $7{,}542$ observations in 2016 takes around $183$ seconds. While a single likelihood evaluation is feasible, since each full Metropolis-Hasting step requires four covariance matrix inversions it would take over five months to compute $20{,}000$ samples using the Cholesky method even when computing each year's likelihood in parallel.

	\subsection{Vecchia Approximation}\label{vecchiaProcess:description}
	
	To render Bayesian estimation feasible we use the Vecchia approximation of \citet{vecchia1988estimation} which approximates a Gaussian process by writing the likelihood as a series of conditional likelihoods and then restricting the set of observations used by the conditional likelihoods. Specifically, let $H_{\textnormal{obs},1},\dots,H_{\textnormal{obs},n_{\textnormal{obs}}}$ denote the set of observations and let $g(1),\dots,g(n_{\textnormal{obs}})$ denote sets of integers such that $g(1)=\varnothing$ and $g(j)\subseteq\{1,\dots,{j-1}\}$. When $g(j)$ is more than one integer, we will let $H_{\textnormal{obs},g(j)}$ denote the set of observations corresponding to the indices in $g(j)$. With this the Vecchia likelihood is defined as
	\begin{equation}
			\ell_{\textnormal{Vecchia}}(H_{\textnormal{obs}})=\sum_{j=1}^{n_{\textnormal{obs}}} \ell(H_{\textnormal{obs},j}|H_{\textnormal{obs},g(j)})\label{vecclik} 
	\end{equation}
	where $ \ell(H_{\textnormal{obs},j}|H_{\textnormal{obs},g(j)})$ is the conditional log-likelihood of $H_{\textnormal{obs},j}$ given $H_{\textnormal{obs},g(j)}$. When $g(j)=\{1,\dots,j-1\}$, each observation is conditioned on all of the observations preceding it in the ordering, and the Vecchia log-likelihood is equivalent to the Gaussian process log-likelihood. When $g(j)$ is a strict subset of the preceding indices, then the Vecchia likelihood is an approximation. The computation of each $\ell(H_{\textnormal{obs},j}|H_{\textnormal{obs},g(j)})$ term requires computing the Cholesky factorization for observations within $g(j)$, so computational gains are achieved when the $g(j)$ are chosen to be small. Generally, the conditioning sets are chosen to be such that $|g(j)|<m$ for some $m<<n_{\textnormal{obs}}$ so that the computation time of the Vecchia likelihood is $O(nm^3)$, as opposed to $O(n^3)$ for the full likelihood. The parameter $m$ is tunable, with larger values yielding a more accurate approximation and smaller values enabling faster computation times. An additional advantage of the Vecchia likelihood is that the terms in Equation \ref{vecclik} can be computed in parallel whereas parallelization cannot be used directly when computing the Cholesky factorization for the full covariance matrix.

	There are three steps for constructing the Vecchia process for a particular set of observation locations. The first step is to order the locations, which is important because the conditioning sets for a particular observation can only include observations that precede it in the ordering. For this step, we use the max-min distance ordering advocated for by \citet{guinness2018permutation}. The second step is to construct the conditioning sets such that each set has no more than $m$ observations. We take the conditioning sets to be the observation's $m$-nearest neighbors under the logic that nearby observations are the most influential and as such will yield the closest approximation to the Cholesky likelihood. The third step is to group the observations whose conditioning sets have substantial overlap. Computing the conditional likelihoods on the grouped observations has been shown by \citet{guinness2018permutation} to increase the accuracy of the approximation while simultaneously decreasing the computation time.  These three steps must be done separately for each year of the Argo data, as the observation locations differ between the years. However, since the conditioning sets and groups do not depend on the parameters, the Vecchia construction needs to only be done once before the beginning of the sampling procedure. 
	
	For a particular ordering, conditioning configuration, and grouping, it can be shown that the Vecchia likelihood induces a valid stochastic process on the domain of observation locations  \citep{datta2016hierarchical}. This "Vecchia process" can be seen as an approximation of the full Gaussian process with more computationally efficient likelihood evaluations. Under this view the Bayesian inference procedure described in Section \ref{modelFramework} can be used with the Vecchia process by simply substituting the full Cholesky likelihood with the Vecchia likelihood. Furthermore, the inverse Cholesky decomposition of the covariance matrix implied by the Vecchia process can be constructed directly as a sparse matrix. To see this we will follow the notation of \citet{katzfuss2021general} and rewrite the summands in Equation \ref{vecclik} as $\ell(H_{\textnormal{obs},j}|H_{\textnormal{obs},g(j)})=\mathcal{N}(H_{\textnormal{obs},j}|B_jH_{\textnormal{obs},g(j)},D_j)$, where $\mathcal{N}$ indicates the normal density, $B_j=k(\textbf{x}_{\textnormal{obs},j},\textbf{x}_{\textnormal{obs},g(j)})k(\textbf{x}_{\textnormal{obs},g(j)},\textbf{x}_{\textnormal{obs},g(j)})^{-1}$, and $D_j=k(\textbf{x}_{\textnormal{obs},j},\textbf{x}_{\textnormal{obs},j})-{B}_jk(\textbf{x}_{\textnormal{obs},g(j)},\textbf{x}_{\textnormal{obs},j})$. Then the elements of the inverse Cholesky factor $\textbf{U}\in\mathbb{R}^{n_{\textnormal{obs}}\times n_{\textnormal{obs}}}$ are $\textbf{U}_{jj}=D_j^{-1/2}$ for all $j$, $\textbf{U}_{ij}=-(B_j)^T_{i,g(j)}D_j^{-1/2}$ for $i\in g(j)$, and $\textbf{U}_{ij}=0$ for $i\notin g(j)$. The sparse matrix $\textbf{U}$ can be used in place of $C^{-1}$ in Equation \ref{lik_chol} to compute the Vecchia likelihood. This is no more computationally intensive than computing the sum in Equation \ref{vecclik} due to $\textbf{U}$'s known sparsity structure determined by the conditioning sets.
	
	It can be seen from the construction of $\textbf{U}$ that for any two observations $j<i$, when $j\notin g(i)$ the partial correlation between $H_{\textnormal{obs},j}$ and $H_{\textnormal{obs},i}$ will be zero, or in other words $H_{\textnormal{obs},j}$ and $H_{\textnormal{obs},i}$ will be conditionally independent. As such the Vecchia process can be seen as an alteration of the full Gaussian process where some conditional dependencies are removed. We note that this does not necessarily mean that $H_{\textnormal{obs},j}$ and $H_{\textnormal{obs},i}$ are independent, as there could be a path of conditional dependencies between observations through the conditioning sets. As such the nearest-neighbors method of constructing the conditioning sets does not simply eliminate dependencies between distant observations, as is done with other likelihood approximation methods such as covariance tapering. This is important to note in our model since we are using the Vecchia process to estimate spatially varying correlation length scales, which could potentially be biased if only dependencies between nearest neighbors were taken into account in the likelihood.

	\subsection{Predictions}\label{vecchiaProcess:predictions}
	
	As discussed in section \ref{modelFramework:posterior}, the estimate of globally integrated heat content is calculated as the area-weighted sum of unobserved ocean heat content values at grid-points along a grid densely partitioning the domain. This is written as $\textnormal{OHC}_\textit{yr}=a_{\textnormal{globe}}^T\textbf{H}_{\textnormal{globe};\textit{yr}}$, where $\textbf{H}_{\textnormal{globe};\textit{yr}}$ are the unknown heat content values at the grid-points and $a_{\textnormal{globe}}$ are the areal weights of the grid-cells. For sufficiently dense grids, computing the joint conditional distribution over the grid-points as in Equation \ref{ohc_post_sd} is computationally infeasible. As such, in order to compute the posterior distribution of heat content, it is essential to be able to efficiently compute the joint posterior distribution for a large number of observations. Fortunately, this can be done by augmenting the Vecchia process described above to include the grid-points at which we will be predicting heat content values. To do this we follow the "observation-first" ordering in \citet{katzfuss2020predictions}, which involves ordering the prediction locations $H_{\textnormal{pred},n_{\textnormal{obs}}+1},\dots,H_{\textnormal{pred},n_{\textnormal{obs}}+n_{\textnormal{grid}}}$, appending them to the ordering of the observation locations, and then computing the nearest-neighbor conditioning sets for the entire set of indexed locations $1,\dots,n_{\textnormal{obs}}+n_{\textnormal{grid}}$. Since conditioning sets can only contain preceding observations, the conditioning sets of the observations remain the same while the conditioning sets of prediction locations can contain any observation location as well as any prediction location preceding it in the ordering.

	Recall that $\textbf{U}$ is the inverse Cholesky factor for the Vecchia process, and let $\textbf{W}=\textbf{U}_{\textnormal{pred},\textnormal{all}}^T\textbf{U}_{\textnormal{all},\textnormal{pred}}$. Then it can be shown that the distribution of the prediction values conditioned on the observations is $\textbf{H}_{\textnormal{pred}}|\textbf{H}_{\textnormal{obs}}=N(-\textbf{W}^{-1}\textbf{U}_{\textnormal{pred},\textnormal{all}}\textbf{U}_{\textnormal{all},\textnormal{obs}}\textbf{H}_{\textnormal{obs}},\textbf{W}^{-1}).$ It follows that the posterior distribution of ocean heat content given the observations is \begin{equation}
			\textnormal{OHC}_\textit{yr}|\textbf{H}_{\textnormal{obs};\textit{yr}}\sim N(-a_{\textnormal{globe}}^T\textbf{W}^{-1}\textbf{U}_{\textnormal{pred},\textnormal{all}}\textbf{U}_{\textnormal{all},\textnormal{obs}}\textbf{H}_{\textnormal{obs};yr},a_{\textnormal{globe}}^T\textbf{W}^{-1}a_{\textnormal{globe}}). \label{ohc_vecchia}
	\end{equation}  As $\textbf{U}$, and subsequently $\textbf{W}$, are sparse, the terms $\mu_{\textnormal{OHC}_\textit{yr}}\equiv-a_{\textnormal{globe}}\textbf{W}^{-1}\textbf{U}_{\textnormal{pred},\textnormal{all}}\textbf{U}_{\textnormal{all},\textnormal{obs}}\textbf{H}_{\textnormal{obs};yr}$ and $\sigma_{\textnormal{OHC}_\textit{yr}}\equiv a_{\textnormal{globe}}^T\textbf{W}^{-1}a_{\textnormal{globe}}$ are both fast to compute. If the full Gaussian process was used to compute this posterior distribution using Equation \ref{ohc_post_sd}, every element of the dense covariance matrix of grid-points would need to be evaluated, which is intractable for sufficiently dense grids.

	\subsection{Evaluating the Accuracy of the Vecchia Process}\label{vecchiaProcess:accuracy}
	
	The accuracy of the Vecchia likelihood as an approximation to the log-likelihood depends on the ordering, conditioning sets, and grouping used in the construction. In particular, the maximum size of the conditioning sets, determined by the parameter $m$, needs to be chosen to balance the trade-off between accuracy and computation time. In order to assess the accuracy of the Vecchia process for our model for different values of $m$, we divided the global ocean into nine basins. These basins are small enough such that the likelihoods and heat content predictions can be computed easily using the full Cholesky decomposition so that the results from the Cholesky and Vecchia methods can be compared. The basins are defined using the latitudinal and longitudinal limits displayed in Figure S1 in the supplementary material. We note that these basins are for evaluating the approximation accuracy only and that the results in Section \ref{results} are computed on the global ocean.

	To evaluate the accuracy of the Vecchia approximation on the ocean heat content field, we fit the model described in Section \ref{modelFramework} independently for each of the nine basins using both the Cholesky method and the Vecchia approximation method for each of $m\in\{10,25,50,100\}$. Using the resulting posterior distributions, the mean, trend, and anomaly fields were interpolated for each basin independently, and globally integrated values were computed using the results within each basin. For the Vecchia process with conditioning parameter $m$ we will refer to the globally integrated values of the mean, trend, and anomaly fields as $\mu_{2007}^{\textnormal{vecc}(m)}$, $\beta^{\textnormal{vecc}(m)}$, and $\textit{anom}_{yr}^{\textnormal{vecc}(m)}$ respectively. Analogously the globally integrated values for the the Cholesky method will be referred to as $\mu_{2007}^{\textnormal{chol}}$, $\beta^{\textnormal{chol}}$, and $\textit{anom}_{yr}^{\textnormal{chol}}$. By separating the mean, trend, and anomaly fields we can examine how close the Vecchia process approximates these three components. We would expect the error in the mean field to be the smallest, as the mean field is highly informed by the values in the data and is not particularly sensitive to the estimated correlation structure. On the other hand, we would expect the anomaly field to have the largest errors, as the predicted anomaly values at unobserved locations are highly sensitive to the correlation structure.

	\begin{table}[ht]
		\centering
		\begin{tabular}{rllr}
			m & mean field & trend field &  anomaly field \\ 
			\hline
			10 & 8.078$\times 10^{-4}$ & .05029 & .7421 \\ 
			25 & 7.764$\times 10^{-5}$ & 1.400$\times 10^{-3}$ & .4081 \\ 
			50 & 2.987$\times 10^{-5}$ & 5.498$\times 10^{-5}$ & 1.665$\times 10^{-3}$  \\ 
			100 & 1.87$\times 10^{-5}$ & 1.963$\times 10^{-5}$ & 1.656$\times 10^{-6}$ \\ 
			\hline
		\end{tabular}
		\vspace{.1in}
		\caption{Fractional errors for the mean, trend, and anomaly fields of the Vecchia approximation as measured against the analogous values computed using the Cholesky method. Global fields were constructed from results obtained from fitting the model to each of the nine basins in Figure S1 independently.}
		\label{basin_analysis}
	\end{table}

	Treating the Cholesky values as "truth", fractional errors $\frac{|\mu_{2007}^{\textnormal{vecc}(m)}-\mu_{2007}^{\textnormal{chol}}|}{\mu_{2007}^{\textnormal{chol}}}$, $\frac{|\beta^{\textnormal{vecc}(m)}-\beta^{\textnormal{chol}}|}{\beta^{\textnormal{chol}}}$, and $\frac{|\textit{anom}_{yr}^{\textnormal{vecc}(m)}-\textit{anom}_{yr}^{\textnormal{chol}}|}{|\textit{anom}_{yr}^{\textnormal{chol}}|}$ are displayed in Table \ref{basin_analysis} for each value of $m$. For the fractional error of the anomaly fields the median over years is displayed. For $m=10$ and $m=25$, the mean field has smaller fractional errors than the trend or anomaly fields, as expected. The accuracy of the mean field does not improve noticeably when increasing $m$ beyond $25$, however the error at that level is acceptably small. The approximation error of the trend and anomaly fields are relatively more substantial at $m=10$, with fractional errors of $5\%$ and $74\%$ respectively.  They improve only marginally when increasing $m$ to $25$ but see a more substantial improvement when increasing to $m=50$, with errors around five one-thousandth of a percent and a tenth of a percent respectively. When further increasing to $m=100$, the approximation errors for the mean and trend fields do not meaningfully change, although the error for the anomaly field does noticeably improve. As the fractional error in the trend field does not decrease after $m=50$, and at $m=50$ the level of error in the anomaly fields is acceptable, we will use the Vecchia process with $m=50$ for the global model fit.

	\section{Results}\label{results}
	
	\subsection{Initial Configuration}\label{results:initConfig}

	As the number of parameters involved in our model is large, it is particularly important to initialize the MCMC sampler at a "good" initial configuration for the sampler to converge in a reasonable number of iterations. To obtain a suitable initial configuration, we used a moving-window estimation method analogous to that used in \citet{kuusela} to obtain values for the spatially varying parameter fields. In this approach, the domain is partitioned according to a grid, and at each grid-point, the observations within a window centered around that point are considered to follow a stationary Gaussian process with process variance, nugget variance, latitudinal range, longitudinal range, mean, and trend parameters. Since each of these windows contains a relatively small number of observations, the locally stationary parameters can be estimated quickly using maximum likelihood estimation with Cholesky factorizations.

	Parameters were estimated using windows of size $20^\circ \times 20^\circ$ centered at each grid-point of a $6^\circ\times 6^\circ$ grid. Note that the $6^\circ\times 6^\circ$ grid used for the moving window parameter estimates is finer than the $8^\circ\times 16^\circ$ resolution of knot points used in the global model; this is done so that we can estimate the hyperparameters of the parameter fields. To estimate the hyperparameters, stationary and isotropic Gaussian processes were fit to the moving window output for each of the parameter fields, obtaining an estimate of the hyper range, mean, and variance for each process. As described in Section \ref{modelFramework}, the correlation length and variance parameters are assumed to follow a log-normal Gaussian distribution. We constrain the hyper-mean of the trend field $\beta$ to be zero to ensure that the estimate is fully informed by the data. Additionally, the mean and trend fields were assumed to have the same correlation length scale parameter. The results are shown in Table \ref{hyperparameters}, where we summarize the hyper-prior distribution using quantiles since the log-normally distributed priors are non-symmetric. The range hyperparameter for each process is shown in the second column as the "effective range", which we define as the cylindrical distance in degrees at which the correlation is equal to $0.05$, and denote $\gamma_{\textnormal{lat}}$ and $\gamma_{\textnormal{lon}}$ for effective latitudinal and longitudinal range respectively.

	\begin{table}[ht]
		\centering
		\begin{tabular}{rlrrrrl}
			\hline
			&  & Effective Range & $q_{.25}$ & $q_{.5}$ & $q_{.75}$  & Units\\ 
			\hline
			& $\gamma_{\textnormal{lat}}$ & 39.97$^\circ$ & 0.89 & 2.50 & 6.98 & Degrees \\ 
			& $\gamma_{\textnormal{lon}}$ & 44.93$^\circ$ & 0.88 & 5.07 & 29.08  & Degrees \\ 
			& $\sigma^2/\phi$ & 36.59$^\circ$ & .0034 & 0.04 & 0.47& Unitless \\ 
			& $\sqrt{\phi}$ & 39.24$^\circ$ & 1.29 & 2.25 & 3.92  & $GJ/m^2$ \\ 
			& $\mu_{2007}$ & 34.88$^\circ$ & 19.70 & 49.24 & 78.78  & $GJ/m^2$ \\ 
			& $\beta$ & 34.88$^\circ$ & -1.53 & 0.00 & 1.53  & $(GJ/m^2)/$year \\ 
			\hline
		\end{tabular}
		\vspace{.1in}
		\caption{Basis field hyperparameters as estimated using maximum likelihood estimation on parameter fields obtained from a $20^\circ\times 20^\circ$ moving window estimation. The range hyperparameter and the correlation length scale fields $\gamma_{\textnormal{lat}}$ and $\gamma_{\textnormal{lon}}$ are reported in terms of effective range, which is the distance at which the correlation is $0.05$. Quantiles are reported due to the non-symmetric nature of the log-normal distribution used for the covariance parameters. The mean of the trend parameter field $\beta$ was manually constrained to be zero to avoid influencing its estimation in the Bayesian procedure. }
		\label{hyperparameters}
	\end{table}
	
	\begin{figure*}
		\centering
		
		\begin{subfigure}{.48\textwidth}
			\centering
			\includegraphics[width=.9\linewidth]{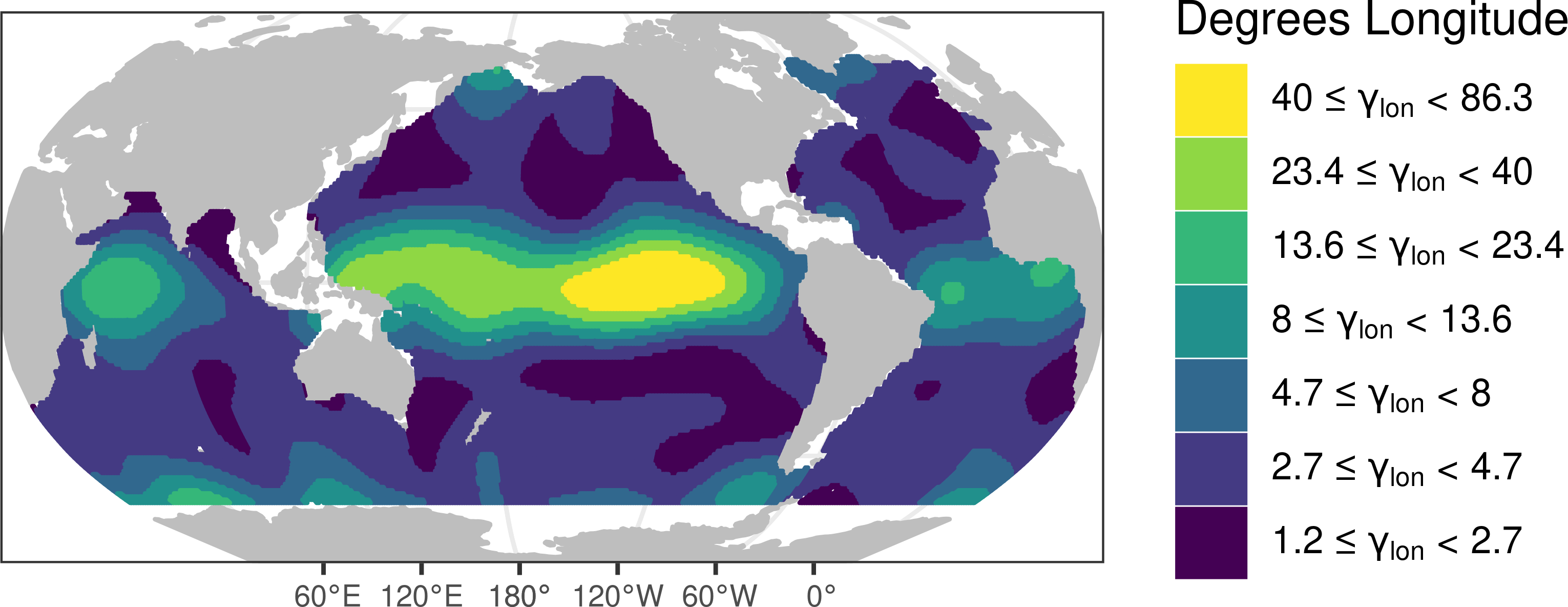}
			\caption{Effective longitudinal range $\gamma_{\textnormal{lon}}$} 	\label{initial_configuration:lon}
		\end{subfigure}	\hspace{.1in}\begin{subfigure}{.48\textwidth}
			\centering
			\includegraphics[width=.9\linewidth]{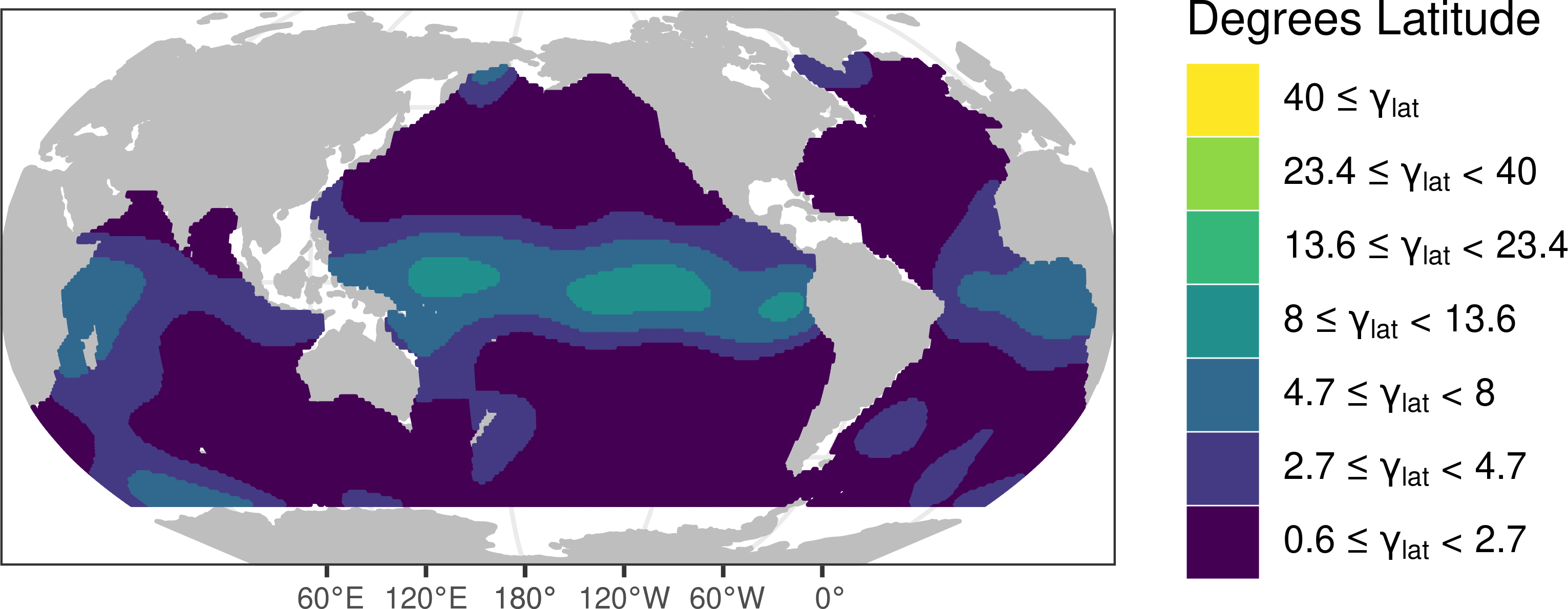}
			\caption{Effective latitudinal range $\gamma_{\textnormal{lat}}$} 	\label{initial_configuration:lat}
		\end{subfigure}
		
		\vspace{.1in}
		
		\begin{subfigure}{.48\textwidth}
			\centering
			\includegraphics[width=.9\linewidth]{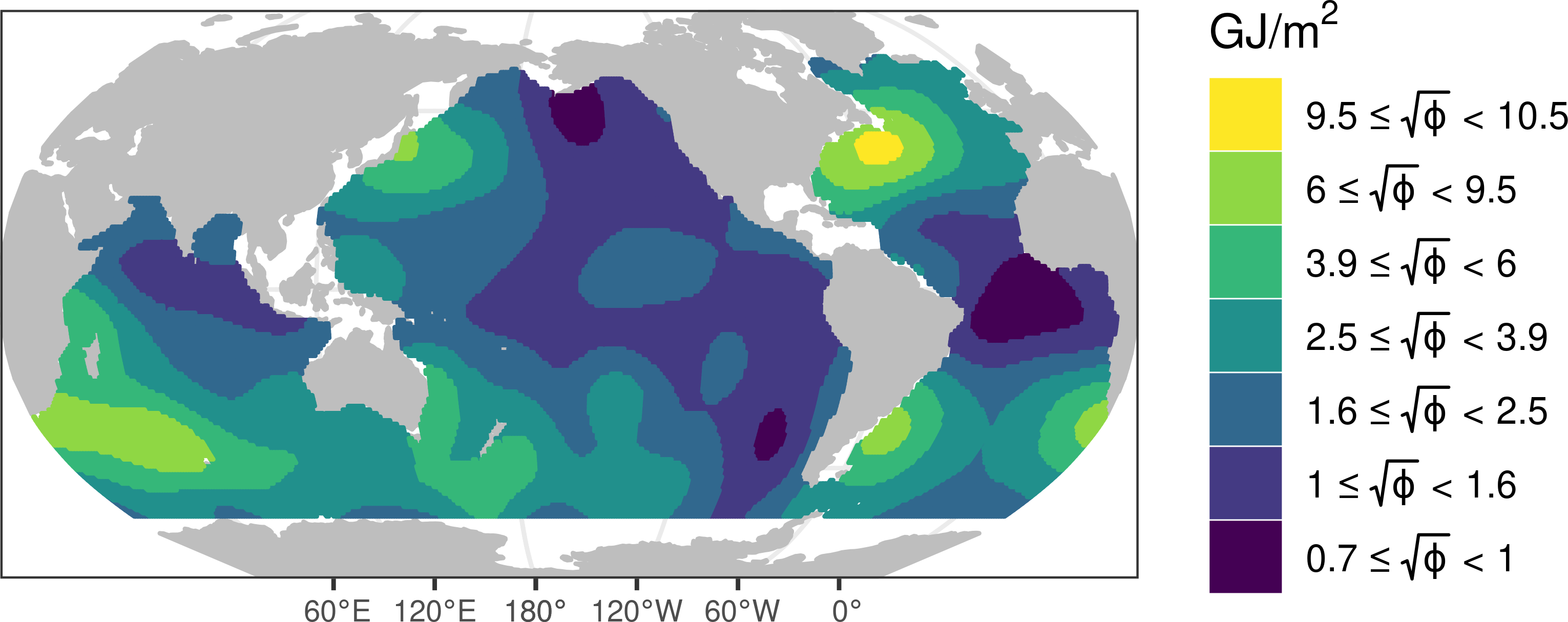}
			\caption{Process standard deviation $\sqrt{\phi}$} 	\label{initial_configuration:phi}
		\end{subfigure}\hspace{.1in}\begin{subfigure}{.49\textwidth}
			\centering  \includegraphics[width=.9\linewidth]{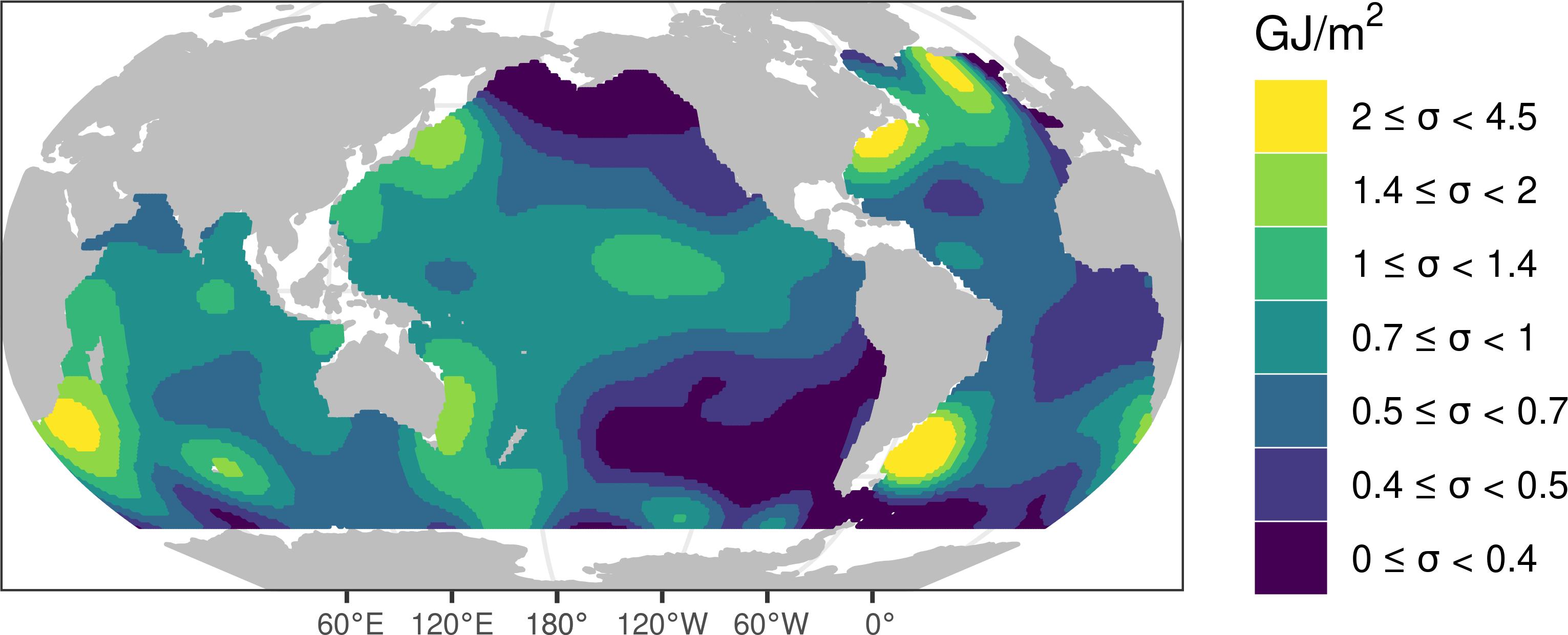}
			\caption{Nugget standard deviation $\sigma$} 	\label{initial_configuration:nugget}
		\end{subfigure}

		\caption{Initial parameter fields obtained from fitting smooth hyperparameter surfaces to the parameter fields obtained from a moving-window approach.} \label{initial_configuration}
	\end{figure*}

	The initial configuration for the MCMC sampler was obtained by kriging the estimated parameters from the moving window approach onto the knot locations using the hyperparameters described in Table \ref{hyperparameters}. Figure \ref{initial_configuration} displays the initial configuration interpolated over a dense grid for display purposes. Similar to Table \ref{hyperparameters}, standard deviations $\sqrt{\phi}$ and effective ranges $\gamma_{\textnormal{lon}}$ and $\gamma_{\textnormal{lat}}$ are shown rather than the actual parameter fields. While this configuration should not be used to quantify global OHC due to the fact that the estimates were derived assuming local stationarity, comparisons to the 2016 anomaly field displayed in Figure \ref{ohc_field:anom_values} yield insights regarding the ocean heat content correlation structure. First of all, we can see from the initial configuration that the correlation length scales around the equatorial regions are much larger than at higher latitudes, and are particularly large in the equatorial Pacific. This phenomenon can be seen in the heat content anomaly field for 2016 through the coherent structures of positive anomalies particularly visible in the eastern equatorial Pacific. These anomalies primarily extend in the longitudinal direction and to a smaller extent in the latitudinal direction.  Anisotropy is reflected in the initial configuration, where we can see that the scale of the effective longitudinal correlation lengths is generally much larger than those in the latitudinal plot, a difference that is particularly stark in the equatorial Pacific. The phenomenon of latitudinal correlation lengths being generally smaller than longitudinal correlation lengths makes sense from a physical perspective, as changes in latitude are associated with changes in climate to a much greater degree than changes in longitude. 
	
	The areas of the ocean that appear to have particularly high variability in the 2016 anomaly field roughly correspond to the areas of high variance in the initial configuration. Four regions in particular stand out in both of these plots; the North Atlantic east of the United States, the Pacific Ocean east of Japan, the South Atlantic east of Argentina, and a large swath of ocean moving towards the south-east from South Africa to South Australia. We note that most of these regions align with known ocean currents, in particular the Gulf Stream of the North Atlantic, the Kuroshio Current off the coast of Japan, the South Atlantic current off of Argentina, and the {Arctic circumpolar current} in the Southern Ocean \citep{colling2001ocean}. This is unsurprising, as the currents generate large anomalies of either positive or negative sign depending on the particular position and behavior of the current at a given time.

	In theory, the choice of initial configuration will not impact the results of the MCMC sampler if a sufficiently long chain is sampled. However, running the sampler on random initial configurations did not converge after $10{,}000$ iterations. Starting from this initial configuration considerably improves the sampler's convergence properties as will be discussed in Section \ref{results:posteriors}. We have also found that it is particularly challenging to sample the parameter fields and the hyperparameters simultaneously due to the fact that the hyperparameters are not directly informed by the data. To avoid this issue, the hyperparameter values obtained from the moving window approach were treated as fixed and not sampled. This can be interpreted as imposing an informative prior which constrains the variability of the parameter fields to reasonable levels. 
	
	\begin{figure}
		
		\begin{subfigure}{\textwidth}
			\centering
			\includegraphics[width=.4\textwidth]{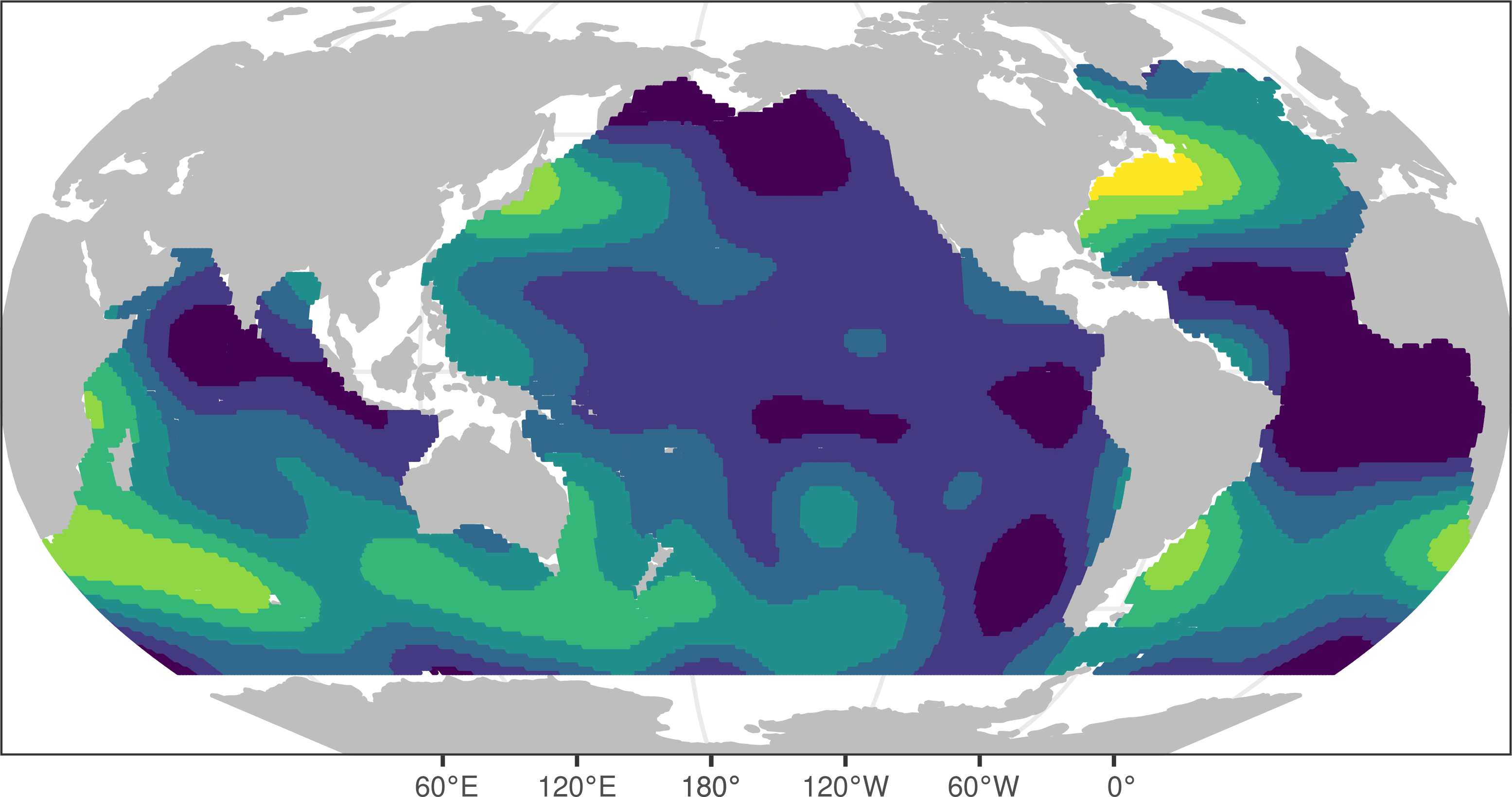}\includegraphics[width=.4\textwidth]{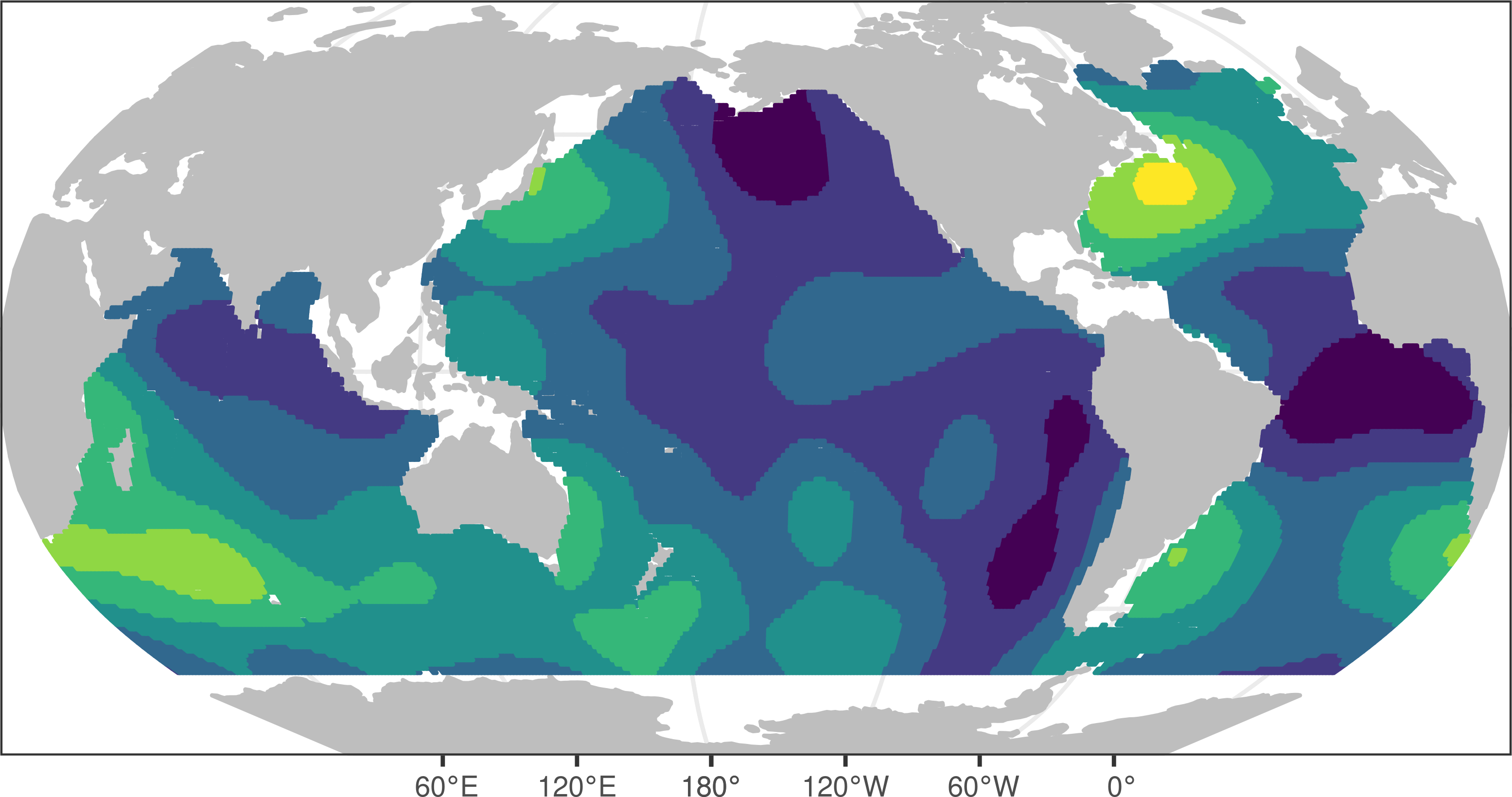} \includegraphics[width=.135\textwidth]{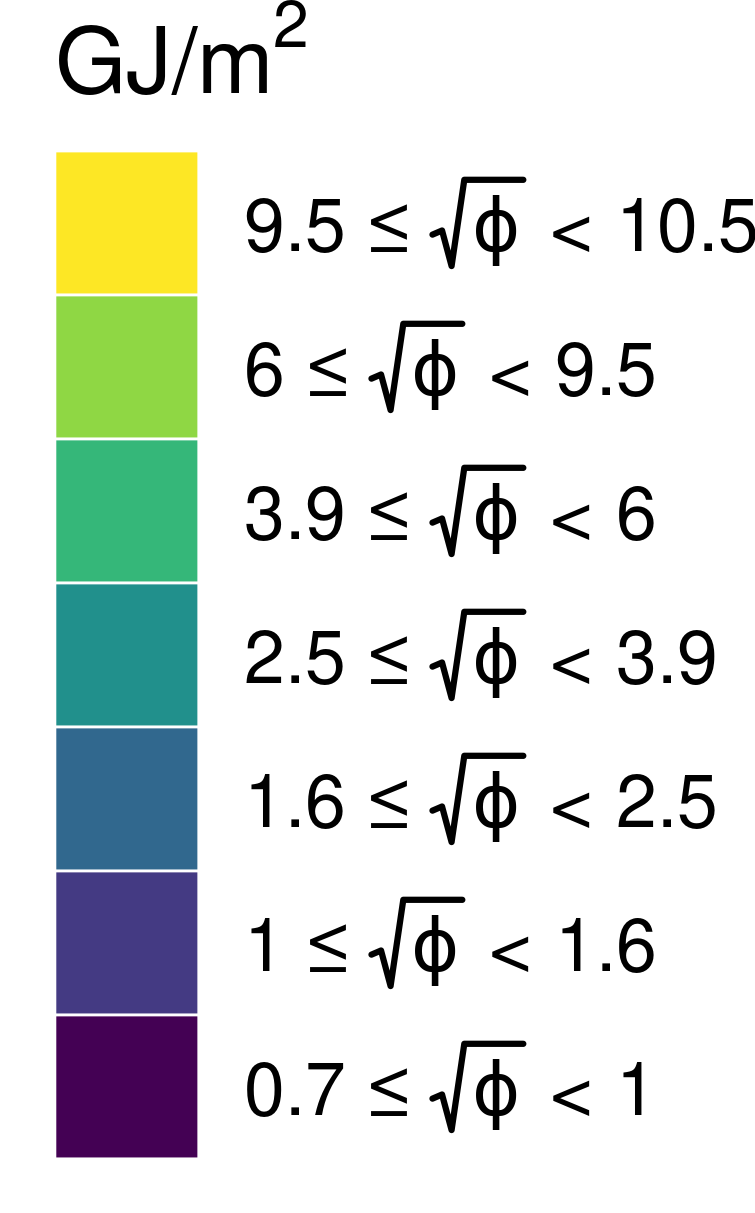}
			\caption{$\sqrt{\phi}$ 5\textsuperscript{th}  and 95\textsuperscript{th} percentile samples} 	\label{post_quantiles:phi}
		\end{subfigure}
		
		\vspace{.1in}
		
		\begin{subfigure}{\textwidth}
			\centering
			\includegraphics[width=.4\textwidth]{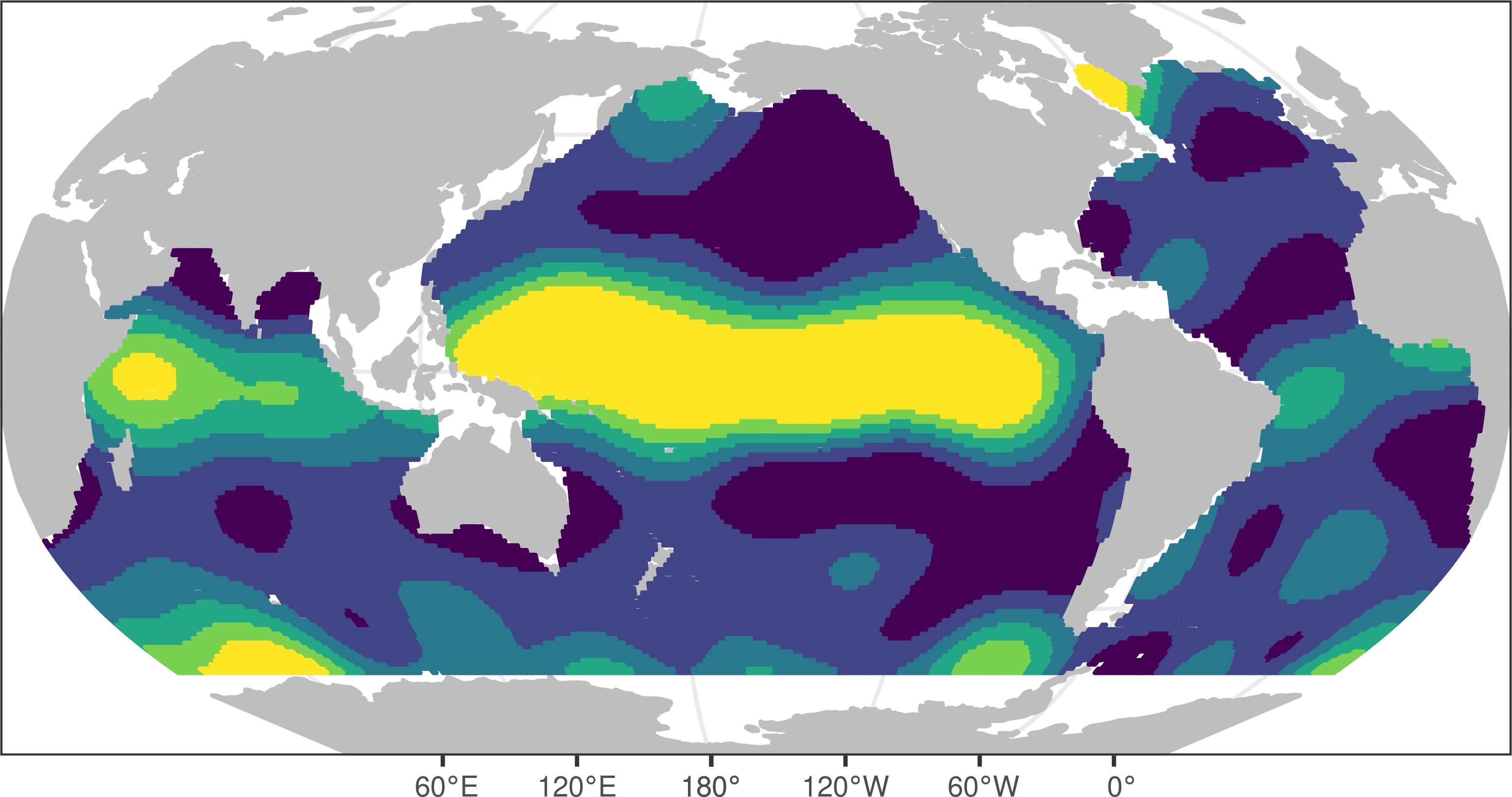}\includegraphics[width=.4\textwidth]{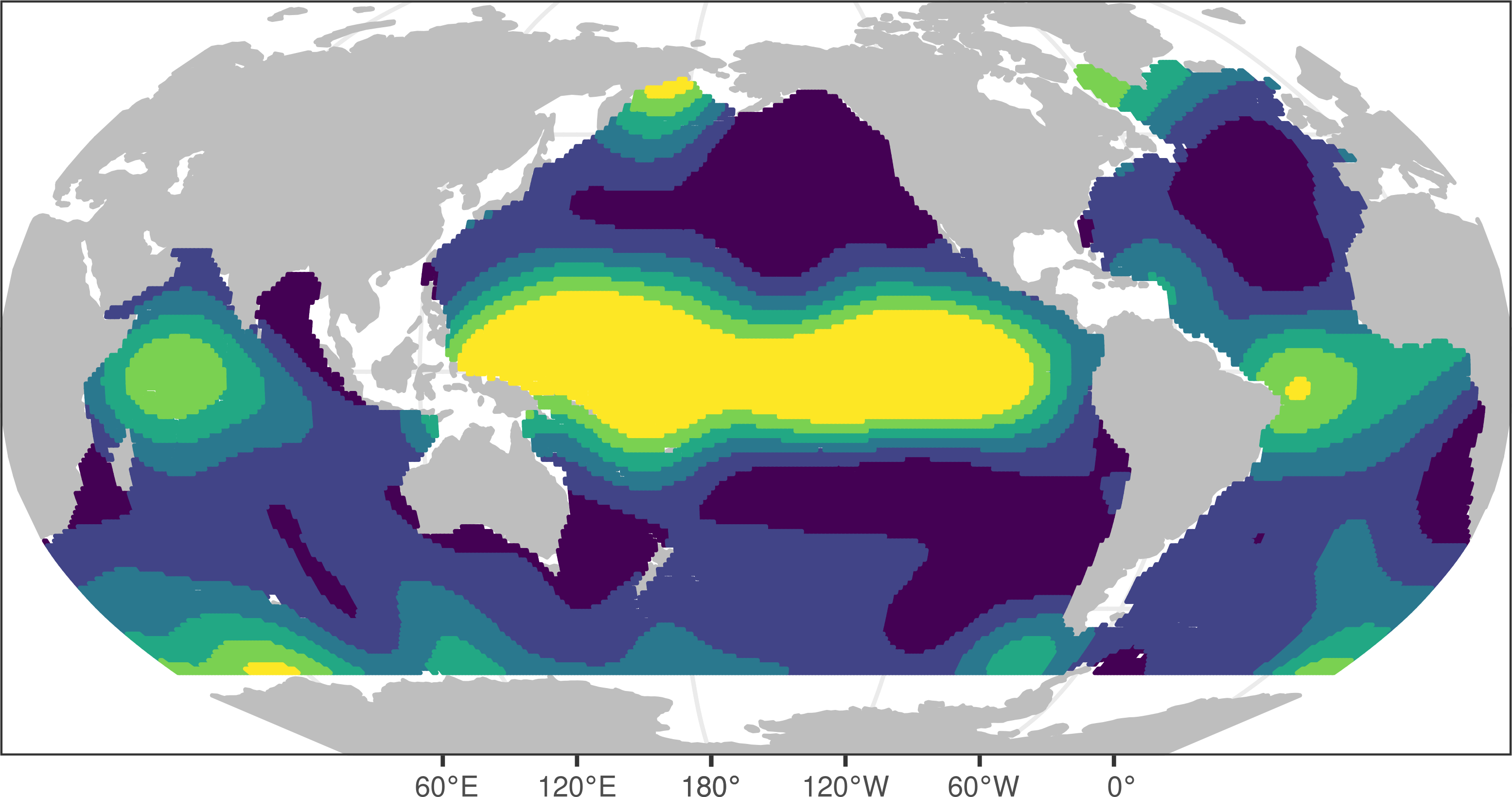}			\includegraphics[width=.135\textwidth]{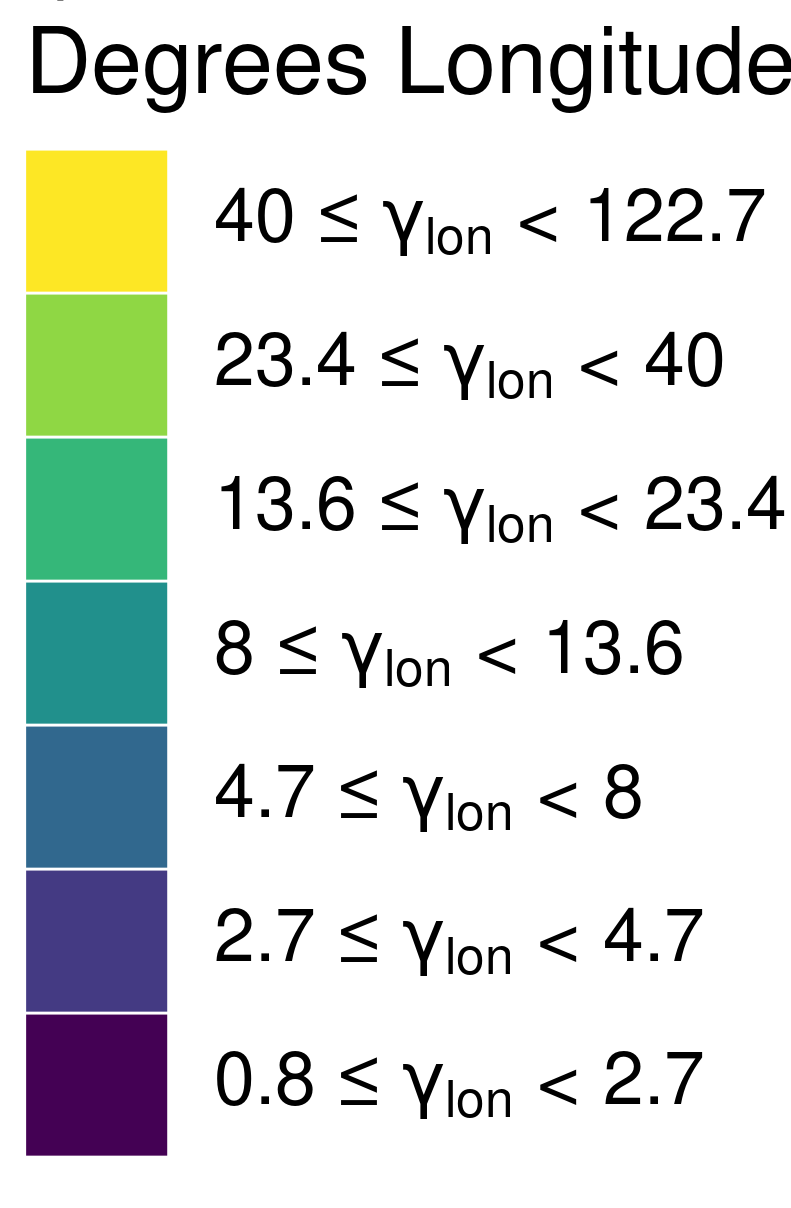}
			\caption{$\gamma_{lon}$  5\textsuperscript{th}  and 95\textsuperscript{th} percentile samples} 	\label{post_quantiles:theta_lon}
		\end{subfigure}
		
		\vspace{.1in}
		
		\begin{subfigure}{\textwidth}
			\centering
			\includegraphics[width=.4\textwidth]{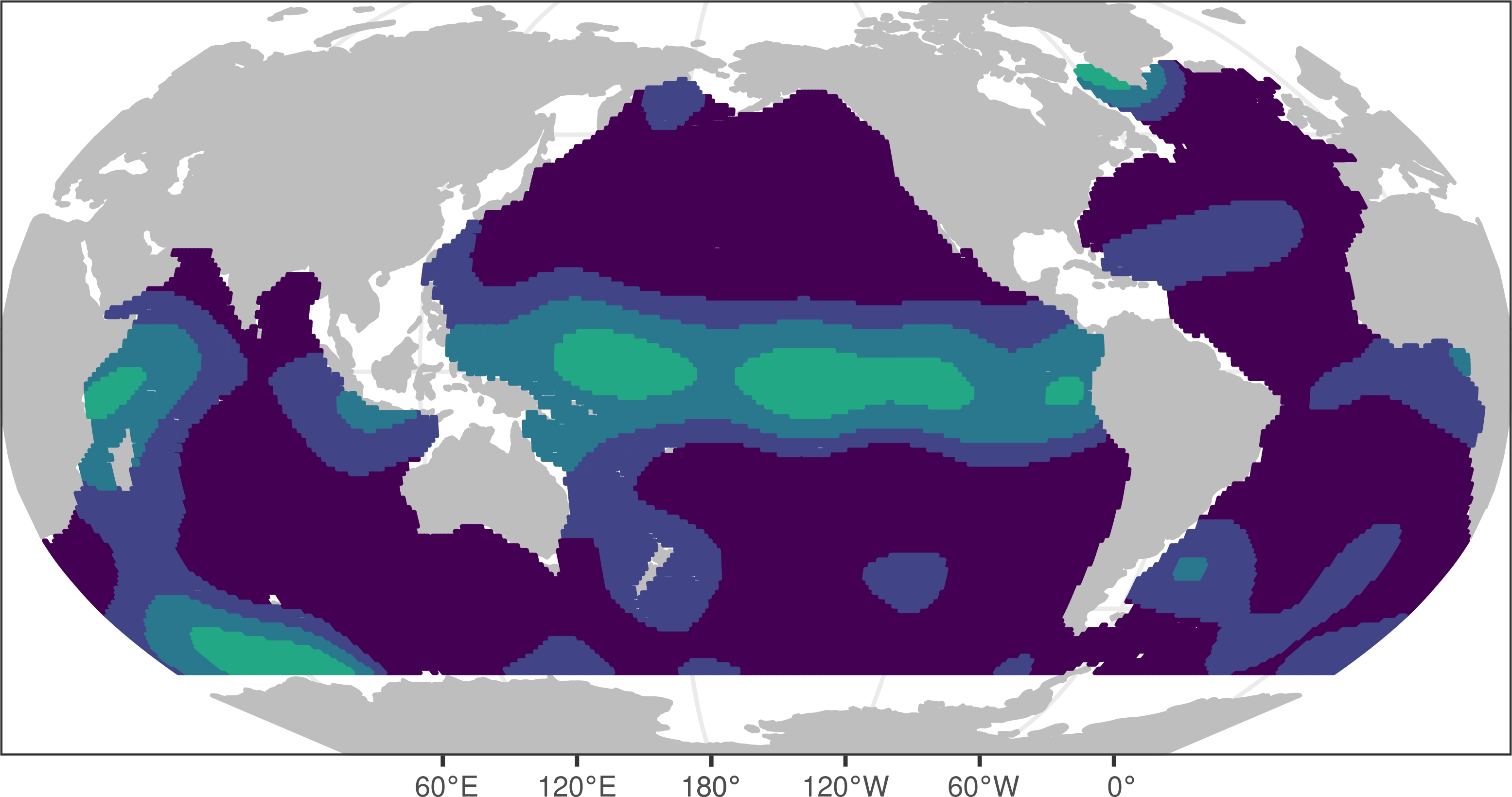}\includegraphics[width=.4\textwidth]{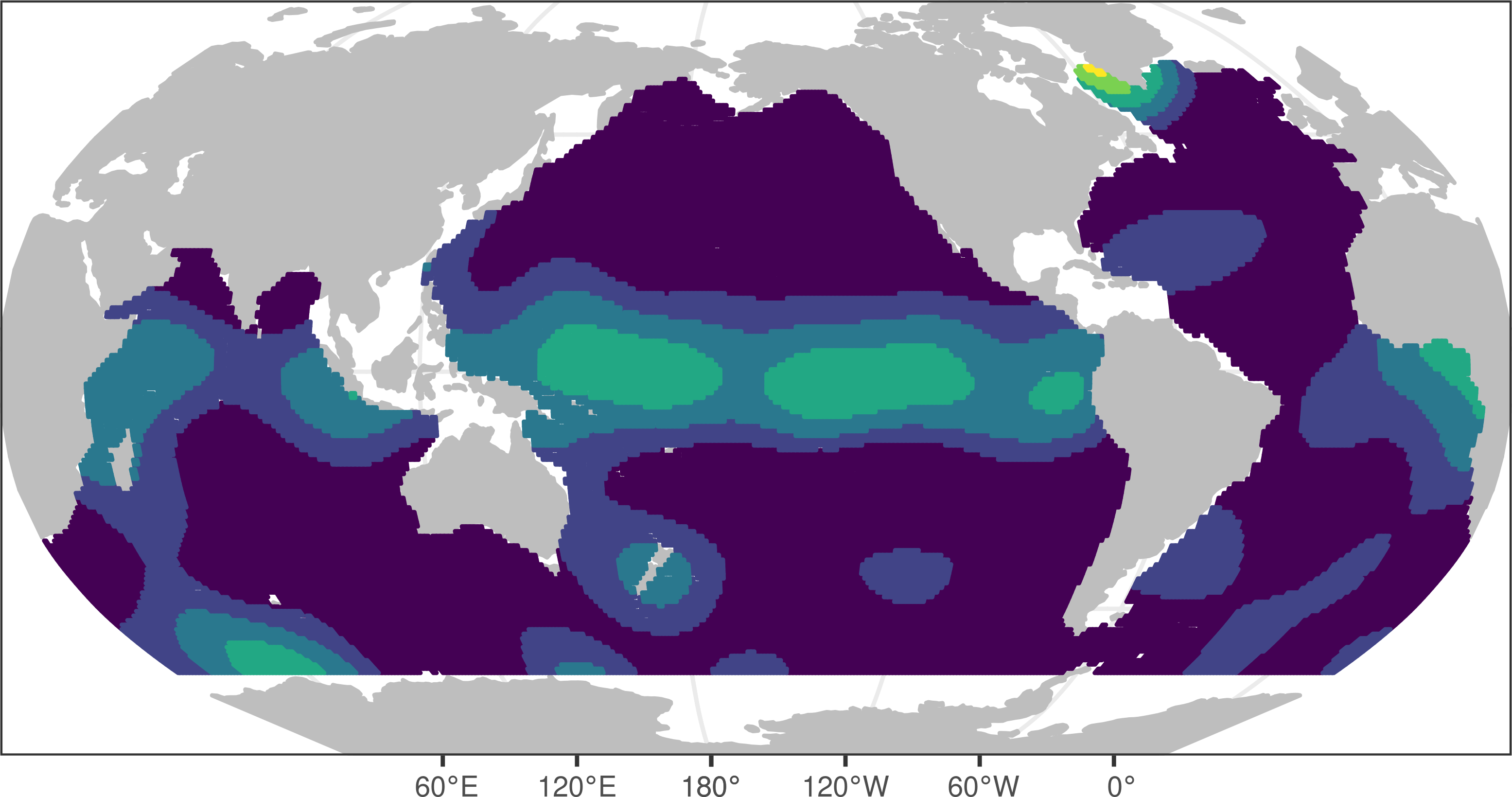}
			\includegraphics[width=.135\textwidth]{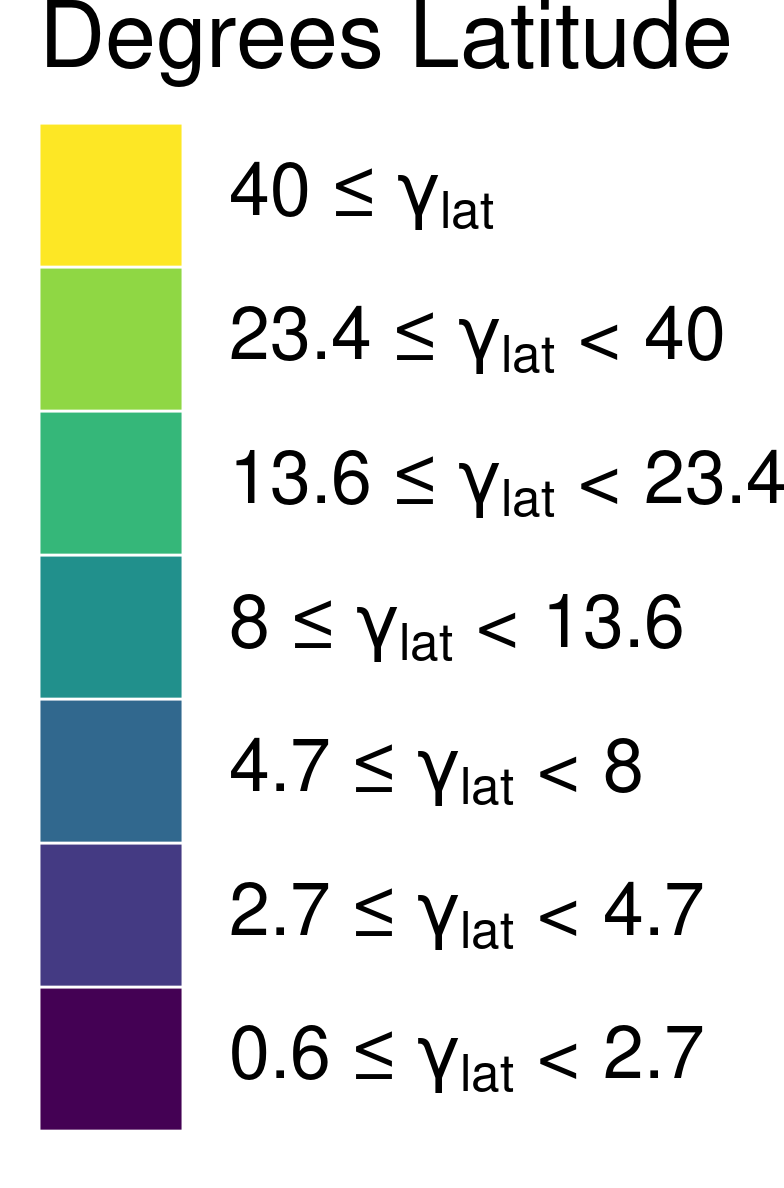}
			\caption{$\gamma_{lat}$  5\textsuperscript{th}  and 95\textsuperscript{th} percentile samples} 		\label{post_quantiles:theta_lat}
		\end{subfigure}
		
		\caption{Samples from the posterior distribution with mean values corresponding to the 5\textsuperscript{th} (left figures) and 95\textsuperscript{th} (right figures) percentiles of the spatial averages of each field from the posterior distribution. }\label{post_quantiles}
	\end{figure}
	
	\subsection{Posterior Distributions}\label{results:posteriors}
	
	Starting from the initial configuration obtained in the previous section, the MCMC sampler was run for $20{,}000$ iterations on a desktop computer with $16GB$ of RAM running Ubuntu 20.04 and using ten cores for the parallel computation of the likelihood evaluations and OHC estimations. The ocean heat content distribution was computed using  Equation \ref{ohc_vecchia} on a $1^\circ\times 1^\circ$ grid every ten iterations. To assess convergence, the  Heidelberger and Welch test \citep{heidelberger1981spectral} was run on the sequence of log posterior densities, yielding convergence after a burn-in period of $5{,}401$ iterations. The results that follow are computed on the posterior samples with the burn-in period removed. 
	
	To get a sense for the posterior distribution of the parameter fields, for each parameter field we computed the average value across space for each sample and identified fields corresponding to the 5\textsuperscript{th} and 95\textsuperscript{th} percentiles of the distribution of mean values. These parameter configurations for the standard deviation $\sqrt{\phi}$, effective latitudinal range $\theta_{\textnormal{lat}}$, and effective longitudinal range $\theta_{\textnormal{lon}}$ are displayed in Figure \ref{post_quantiles}. Note that these are not the point-wise 5th and 95th percentiles of the parameter estimates, which would not retain the spatial structure present in each sample, but rather spatially coherent members of the posterior distribution. The most noticeable difference between these samples and the initial configuration displayed in Figure \ref{initial_configuration} are the higher longitudinal range parameters in the samples, particularly noticeable in the equatorial West Pacific. This could be due to the fact that the moving-window parameter method may underestimate high values of the range parameter due to the fact that its estimates are based on data from only a $20^\circ\times 20^\circ$ window. Otherwise, the samples appear to be largely similar to the initial parameter fields, further highlighting the benefit of starting the MCMC sampler at a plausible initial configuration. 
	
	\begin{figure}
		\centering
		\includegraphics[width=.7\textwidth]{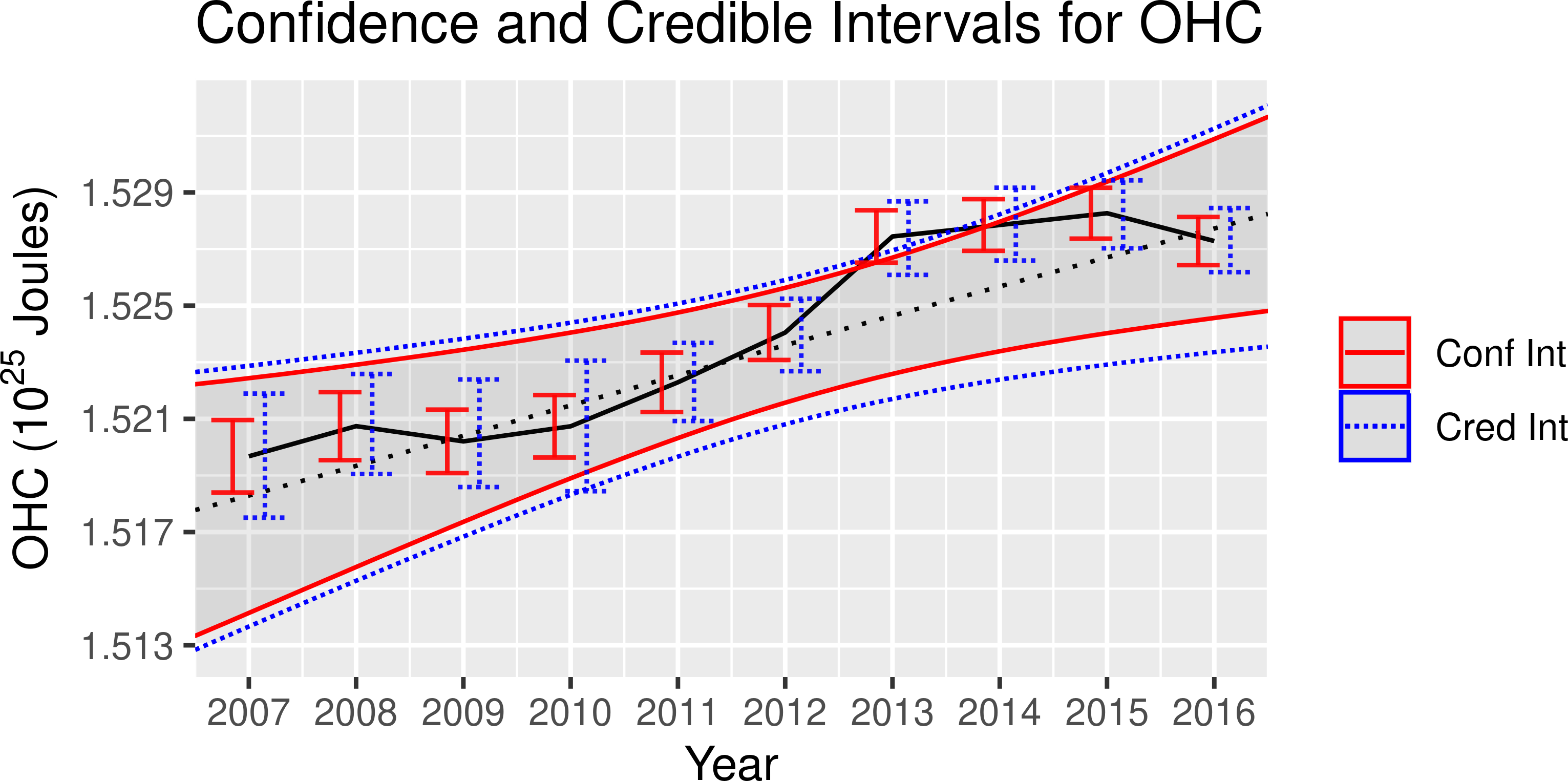}
		
		\caption{Estimation of uncertainty in the ocean heat content values for January of each year and the increasing trend. Confidence intervals in solid red correspond to uncertainty arising from the variability of the heat content field. Credible intervals in dotted blue include the additional uncertainty incurred from parameter estimation. } \label{ohc_by_year}
	\end{figure}

	While the 5\textsuperscript{th} and 95\textsuperscript{th} percentile configurations in Figure \ref{post_quantiles} are fairly similar to each other for each of the parameter fields, the subtle differences between the percentile maps show that there are regions with higher posterior variability. In the process standard deviation maps in Figure \ref{post_quantiles:phi}, differences can be seen in the value and areal extent of the several highly-variable current regions discussed before, indicating some uncertainty in the marginal standard deviation values in these areas. For the longitudinal range maps in Figure \ref{post_quantiles:theta_lon}, variability can be seen in the higher-ranged area in the western Pacific, which is the region that changes the most between the initial configuration and the posterior samples. We can also see variability in the distribution of longitudinal ranges in the Southern Ocean, which is likely due to the fact that parameters in this region are more difficult to estimate due to the smaller number of observations near the extent of Antarctic sea ice, as well as the large variability in the data. For the posterior maps for latitudinal range in Figure \ref{post_quantiles:theta_lat}, we can see variability in the value and extent of the high-ranged area in the equatorial regions, as well as in smaller areas such as south-east of Argentina and around New Zealand. Overall, however, the variability in the posterior distribution appears to be small for each of the parameter fields, indicating that the uncertainty in the ocean heat content and trend estimations induced by uncertainty in the estimation of the parameters will be relatively low.
	
	The ocean heat content mean and variance were calculated for each year for every ten samples after the burn-in period. For each year, we found the median OHC value over the posterior samples and calculated the 95\% confidence interval $\mu_{\textnormal{OHC}_\textit{yr}}^{\textnormal{median}}\pm 1.96\sigma_{\textnormal{OHC}_\textit{yr}}^{\textnormal{median}}$; these are shown as solid red bars in Figure \ref{ohc_by_year}. This represents the uncertainty range induced by the interpolation of the observations to unobserved locations and can be interpreted as a frequentist confidence interval as it does not account for the priors on the parameters. To compute the Bayesian credible interval for each year,  for each iteration $t$ of the sampler  $100$ heat content values were re-sampled from the marginal OHC distribution $\mathcal{N}(\mu_{\textnormal{OHC}_\textit{yr}}^{(t)},\sigma_{\textnormal{OHC}_\textit{yr}}^{(t)})$. Then 5\textsuperscript{th} and 95\textsuperscript{th} percentiles were computed from the pooled re-resampled OHC values.  The re-sampling step ensures that our credible intervals accurately take into account both the interpolation uncertainty from $\sigma_{\textnormal{OHC}_\textit{yr}}^{(t)}$ as well as the parameter uncertainty from the variability of $\mu_{\textnormal{OHC}_\textit{yr}}^{(t)}$ over the samples. These intervals are displayed as dotted blue bars in Figure \ref{ohc_by_year}. We can see that both of the intervals are wider in earlier years, which is to be expected since there are fewer data points in those years. The credible intervals are meaningfully wider than the confidence intervals in earlier years, although the difference becomes small in later years, which is to be expected since the parameter values have a greater effect on the interpolation when there are larger gaps in the data. An analogous procedure was done to obtain uncertainty intervals for the estimated trend; for this parameter, we re-sampled the entire trend field using the posterior covariance matrix conditioned on the other parameters at each iteration. To obtain the 95\% credible interval we identified the 2.5\% and 97.5\% percentiles of the integrated trend values over the re-sampled fields. The 95\% confidence interval, computed with the configuration corresponding to the median integrated trend value, is displayed with solid red lines in Figure \ref{ohc_by_year}, while the 95\% credible interval is displayed as dotted blue lines.  The globally integrated trend is positive and highly significant, with a p-value of $.00165$ obtained from the credible interval.  
	
	\begin{figure}
		\centering
		\begin{subfigure}{.46\textwidth}
			\centering
			\includegraphics[width=1\linewidth]{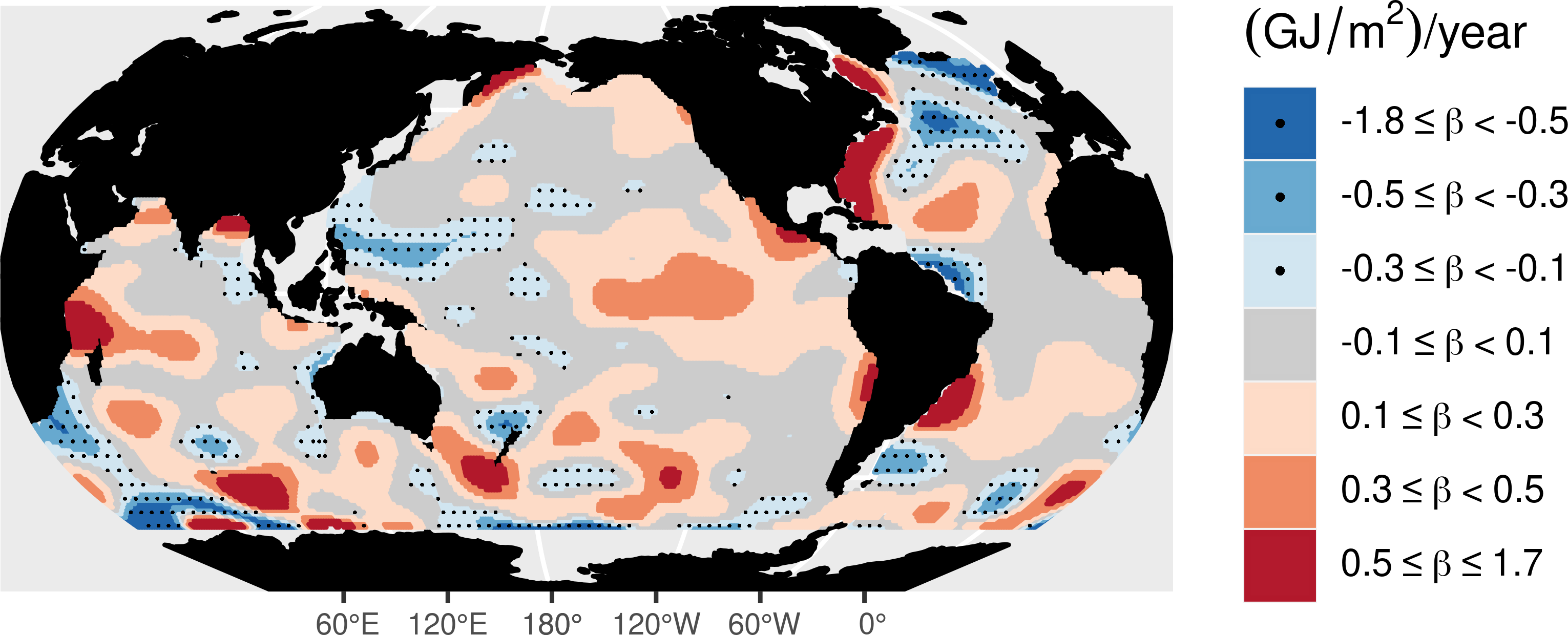}
			\caption{5\textsuperscript{th} percentile trend field} 	\label{slope_posterior:q05}
		\end{subfigure}\hspace{.25in}\begin{subfigure}{.46\textwidth}
			\centering
			\includegraphics[width=1\linewidth]{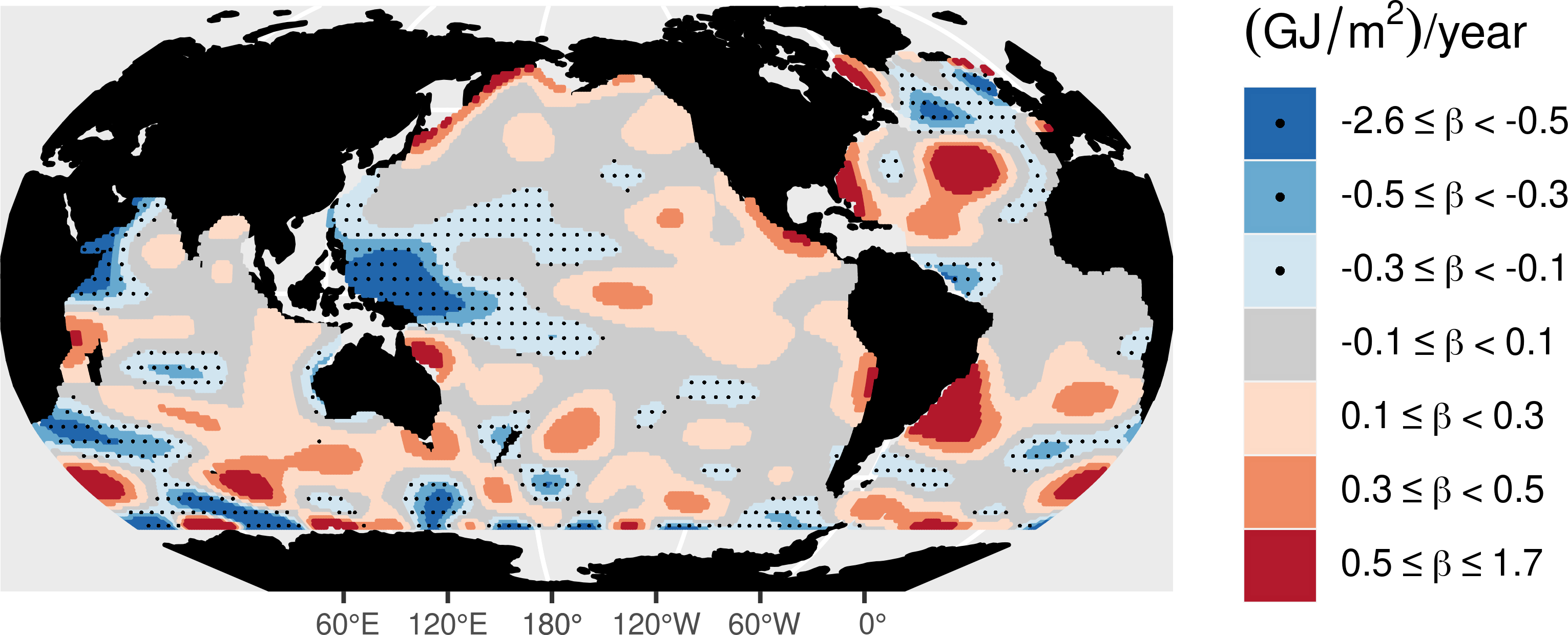}
			\caption{95\textsuperscript{th} percentile trend field} 	\label{slope_posterior:q95}
		\end{subfigure}
		
		\vspace{.1in}
		
		\begin{subfigure}{.5\textwidth}
			\centering
			\includegraphics[width=1\linewidth]{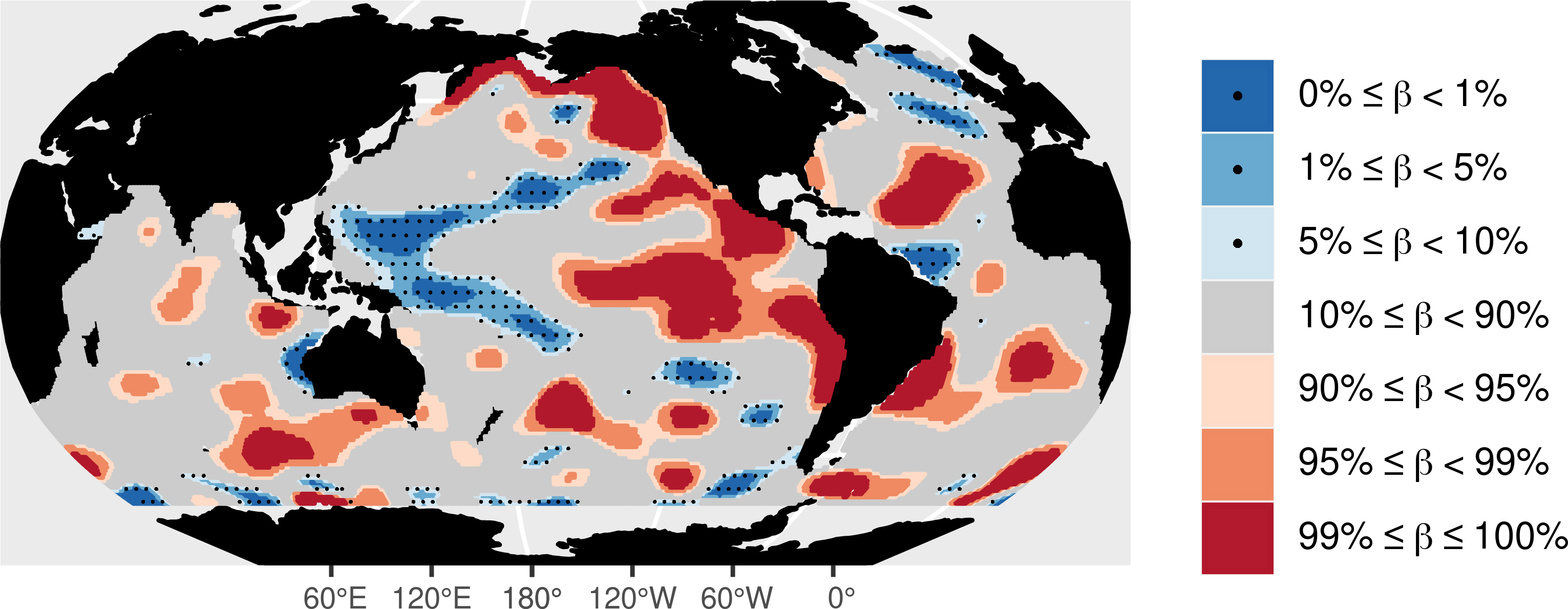}
			\caption{Posterior probability of positive trend} 	\label{slope_posterior:agreement}
		\end{subfigure}
		
		\caption{Sub-figures (a) and (b) show samples of the trend posterior whose integrated values are the 5\textsuperscript{th} and 95\textsuperscript{th} values of the posterior distribution, which are $4.660\times10^{21}$J/year and $16.032\times10^{21}$J/year respectively. Sub-figure (c) gives the posterior probability that an ocean pixel is gaining heat; values less than $10\%$ indicate confidence that the trend is negative.} \label{slope_posterior}
	\end{figure}
	
	In re-sampling not just the globally integrated trend values but the entire field, we are able to obtain additional information about the spatial variability in the ocean heat content trend. Figure \ref{slope_posterior} displays samples of the trend fields whose globally integrated values equal the $5$th (panel a) and $95$th (panel b) percentiles of the re-sampled distribution. We can see that variability in the trend field is higher than that in the correlation and variance parameters in Figure \ref{post_quantiles}. Notable features include the cooling trend in the western Pacific and the warming trend in the eastern Pacific, although there appears to be variability in the intensity and areal extent of these trends. There is more confidence in a warming trend in the ocean south-east of Brazil with little apparent variation in the intensity or spatial extent between the two percentile maps. The North Atlantic, on the other hand, displays variable spatial patterns, with the $5$th percentile field showing a negative trend southeast of the United States that is not present in the $95$th percentile map, and while the two percentile fields agree on a cooling trend in the far North Atlantic they differ on its intensity. Other regions where the trend pattern appears largely uncertain are in the Indian Ocean and the Southern Ocean. 
	
	These two percentile maps can only give an approximate view of the variability in the trend field, as there is a high-dimensional space of fields whose globally integrated values lie at the $5$th and $95$th percentiles of the posterior distribution. To obtain a better idea of agreement in the posterior we calculated the percentage of the re-sampled trend fields that agree on the sign of the trend for each grid-cell of a $1^\circ\times 1^\circ$ grid. This is equivalent to the posterior probability of a positive trend value, or $P(\beta(x)>0|\textnormal{observations})$, and the results are displayed in Figure \ref{slope_posterior:agreement}. We can see that there is agreement on certain spatial features such as the western Pacific cooling trend and the eastern Pacific warming trend. As expected, there is high-level agreement on the warming trend in the ocean south-east of Brazil, and while there is agreement on a positive trend region in the mid-latitude North Atlantic and on the cooling trend in the high-latitude North Atlantic there is no agreement on the trend in the area of the ocean south-east of the United States. There is not any strong agreement on general trends in the Indian and Southern Oceans, however, there are isolated bubbles of confident warming and cooling scattered throughout these regions. We do not expect that these small-scale features would remain significant if more years of observations were used to fit the model, which is reserved for future work. 
	
	\section{Model Validation}\label{modelValidation}

	To validate our model we use leave-one-float-out (LOFO) cross-validation which, for each float, removes the observations corresponding to that float and then uses the remaining locations to predict the values at the removed locations. The motivation for LOFO validation is that, in the Argo data, the distance between the three profiles observed by a single float during a month is generally less than that between observations from different floats.  By removing the entire float's worth of observations we are better able to see how well the model is doing at estimating heat content at unobserved areas of the ocean. One downside to this approach is that the spatial area from which the floats are removed will be variable due to the fact that floats are not equally distributed around the ocean. The cross-validation performance was evaluated using mean absolute error (MAE), root-mean-squared error (RMSE), and continuous ranked probability scores (CRPS) \citep{gneiting2005calibrated}. While the MAE and RMSE metrics assess the accuracy of the predictions, the CRPS metric evaluates precision as well by penalizing higher standard errors in the predictions. 
	
\begin{figure}
		\centering
		\includegraphics[width=.7\linewidth]{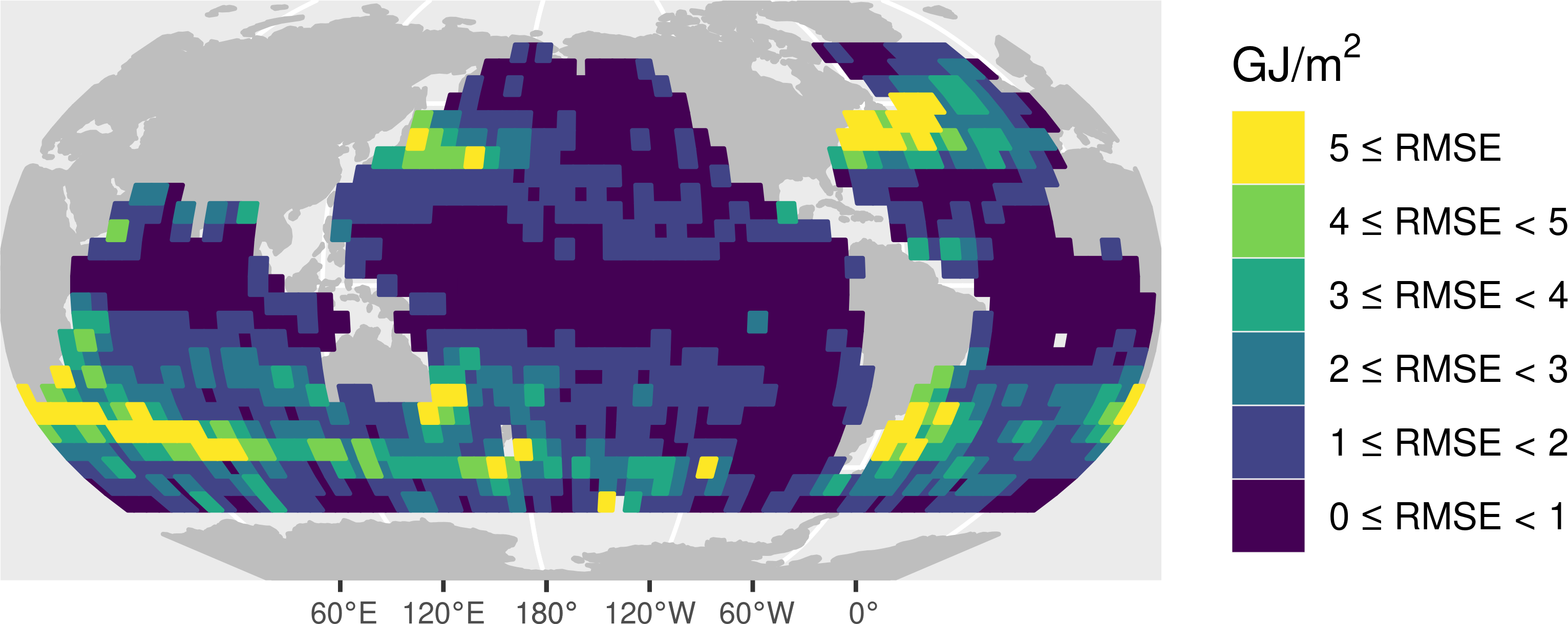}
		
		\caption{Average LOFO MAE over all years for each grid-cell of a $5^\circ\times 5^\circ$ grid.} \label{lofo_grid}
	\end{figure}

	These cross-validation methods were evaluated using the maximum posterior density configuration from the global sampler, and Figure \ref{lofo_grid} shows the LOFO MAE scores averaged over observations from all years within each cell of a $5^\circ\times 5^\circ$ grid.   The globally-averaged LOFO scores for the full model are displayed in the first row of Table \ref{lofo_cv}. We also performed a windowed cross-validation approach as described in Section S4 in the supplementary material, with results showing patterns similar to the LOFO results displayed in Figure S3 and Table S2. We can see from Figure \ref{lofo_grid} that while the validation errors are low in the equatorial regions where the variance is low and the spatial correlation is high, the model performs worse in the high variance and low spatially correlated areas corresponding to the currents. 
	
	Due to the variability of these regions, there is a limit to the predictive ability of any model, and as such the raw validation scores are difficult to put into context without a reference model for comparison. To compare our model to a commonly cited approach in the climate community we have implemented the non-stochastic approach of \citet{levitus2012world} to compute predictions and associated uncertainties in the cross-validation procedures. We have also used the \citet{levitus2012world} approach to compare predicted ocean heat content values and trends, with results presented in Figure S2 and Table S1 in the supplementary material. As we can see in Table \ref{lofo_cv}, our fully non-stationary model produces a $11.2\%$ improvement over the \citet{levitus2012world} approach in terms of RMSE in the LOFO validation. Following the logic of \citet{kuusela}, assuming a $1/\sqrt{n_{\textnormal{obs}}}$ rate of convergence for the Gaussian process predictions, these percent improvements can be seen as equivalent to a $24\%$ increase in the number of observations, or an addition of around $650$ floats.

	\begin{table}[ht]
		\centering
	\begin{tabular}{rllll}
		\hline
		& Package & MAE & RMSE & CRPS \\ 
		\hline
		Fully Non-stationary & BayesianOHC & 1.610 (9.34\%) & 2.708 (11.19\%) & 1.330 (8.15\%) \\ 
		Stationary $\sigma^2/\phi$ & BayesianOHC & 1.612 (9.24\%) & 2.699 (11.47\%) & 1.324 (8.56\%) \\ 
		Non-stationary Isotropic & BayesianOHC & 1.625 (8.50\%) & 2.731 (10.43\%) & 1.344 (7.22\%) \\ 
		Stationary $\theta_{\textnormal{lat}}$ & BayesianOHC & 1.633 (8.04\%) & 2.738 (10.20\%) & 1.347 (7.03\%) \\ 
		Stationary $\theta_{\textnormal{lon}}$ & BayesianOHC & 1.655 (6.81\%) & 2.762 (9.42\%) & 1.364 (5.79\%) \\ 
		Stationary $\phi$ & BayesianOHC & 1.667 (6.12\%) & 2.701 (11.43\%) & 1.364 (5.85\%) \\ 
		Stationary Spatiotemporal & GpGp & 1.705 (3.98\%) & 2.850 (6.53\%) & 1.603 (-10.66\%) \\ 
		Stationary Isotropic & GpGp & 1.718 (3.26\%) & 2.865 (6.03\%) & 1.618 (-11.74\%) \\ 
		Fully Stationary & BayesianOHC & 1.742 (1.94\%) & 2.758 (9.54\%) & 1.431 (1.19\%) \\ 
		Levitus et al. & BayesianOHC & 1.776 (0.00\%) & 3.049 (0.00\%) & 1.448 (0.00\%) \\ 
		\hline
	\end{tabular}	
		\caption{Leave-one-float-out cross-validation scores for the fully non-stationary anisotropic model, restrictions to stationarity in each of the parameter fields, isotropic models, and a spatio-temporal model. Percentage improvements over the reference model \citep{levitus2012world} are displayed in parentheses. Units are GJ/m$^2$.}
		\label{lofo_cv}
	\end{table}

	While it is apparent that our model outperforms the reference model in cross-validation, we are also interested in which non-stationary aspects of our model contribute the most to the improvement. To assess this, for each of the correlation and variance parameter fields, we re-fit the model under the constraint that the parameter field is spatially constant. These models were fit by running the \pkg{BayesianOHC} sampler with the stationarity constraint imposed. Initial configurations were obtained by adjusting the last sample of the full MCMC chain so that the newly stationary field has the median value of the non-stationary field.

	The cross-validation scores with relative improvements over the reference model are shown in Table \ref{lofo_cv}. In terms of mean absolute error, the constrained models have improved cross-validation scores over the fully non-stationary model, indicating the importance of accounting for spatial structure in the model parameters for the ocean heat content field. However, the stationary inverse signal-to-noise and marginal variance models perform slightly better than the full model in terms of RMSE scores, with the stationary signal-to-noise model also having a slightly improved CRPS score. This is likely due to the fact that the fully non-stationary model has higher average values of these parameters than the corresponding stationary models, with the stationary signal-to-noise model having a $\sigma^2/\phi$ value of $.18$ versus the average full model value of $.24$, and the stationary variance model having a $\sqrt{\phi}$ value of $2.28$ versus the average full model value of $2.41$. Unlike MAE, the RMSE and CRPS values take into account error variance and prediction uncertainty, which can explain why the stationary models perform slightly better with regard to those metrics. We note however that non-stationary modeling of the variance parameters is still desirable, as keeping those parameters stationary will likely underestimate the uncertainty in highly variable regions. Increases in all three cross-validation metrics are observed when $\theta_{\textnormal{lat}}$ and then $\theta_{\textnormal{lon}}$ are forced to be stationary, which is as expected given that these parameters, especially $\theta_{\textnormal{lon}}$, vary substantially over the domain (Figure \ref{post_quantiles}). 
	
	Our approach involves the use of a cylindrical approximation for the sphere in order to allow for separate correlation length scales in the latitudinal and longitudinal directions, which is important given the anisotropic behavior of ocean heat content. It is common in the statistics literature to instead treat the sphere as embedded in $\mathbb{R}^3$ space and use the euclidean chord-length distance within an isotropic framework. There are several existing packages including \pkg{GpGp} \citep{guinness2018gpgp}, \pkg{GPVecchia} \citep{katzfuss2020gpvecchia}, and \pkg{BayesNSGP} \citep{turek_bayesnsgp_2019} that allow for isotropic Gaussian process modeling using this metric. While there is not a natural way to represent anisotropy using correlation length-scales with $\mathbb{R}^3$ coordinates, we are interested in examining how well these approaches perform in comparison to ours at cross-validation. To compare against a stationary model we used the "fit\_model" function from the \pkg{GpGp} package with the "matern\_sphere" covariance option to fit an isotropic Matérn model to the data in each year independently. As this package does not immediately allow for a Gaussian process mean field as used in our model, we instead use a mean field parameterization consisting of linear and quadratic terms of latitude and longitude. The results of this model fit are displayed in the "Stationary Isotropic" row of Table \ref{lofo_cv}. While this model has an improved MAE score over the fully stationary anisotropic cylindrical fit, it performs worse in all metrics than the non-stationary models fit using \pkg{BayesianOHC}. As \cite{guinness2021gaussian} have noted that isotropic spatio-temporal modeling yields a higher likelihood than non-stationary spatial-only models, we additionally used the \pkg{GpGp} package to fit a stationary spatio-temporal model to the data, shown in the "Stationary Spatiotemporal" row of Table \ref{lofo_cv}. While this performs better than a spatial-only isotropic model, it still underperforms the non-stationary cylindrical models.  
	
	To evaluate the difference between the isotropic and anisotropic approaches in the non-stationary setting, we used the \pkg{BayesianOHC} package to fit a non-stationary isotropic model using the chord-length distance to the full set of Argo data, with the LOFO cross-validation results displayed in the third row of Table \ref{lofo_cv}. We can see that while the non-stationary isotropic model outperforms most of the other models, the full model achieves superior cross-validation scores. As the only difference between these two models is the cylindrical anisotropy component, this improvement shows the value of representing latitude-longitude anisotropy when modeling ocean heat content.

	\section{Conclusion}\label{conclusion}

	Our kernel-convolution hierarchical Bayesian Gaussian process model for ocean heat content is able to provide uncertainty quantification in ocean heat content values, as well as in the heat content trend, that could not be produced using previous methods. The Gaussian process approach allows us to quantify the uncertainty induced from having to interpolate the ocean heat content field over the domain using information from limited observations.  The kernel-convolution covariance structure allows us to capture the non-stationary covariance properties of the ocean heat content field through spatially varying parameter fields. These convolutions are made possible by using the cylindrical distance, an acceptable distance metric for our application because the model does not extend to the poles and the longitudinal correlation length scales can vary with the sphere's changing radius over latitude. The cylindrical distance metric is a novel approach to representing anisotropy on the sphere that is not possible using an $\mathbb{R}^3$ representation. We have shown that our resulting fitted covariance model achieves validation scores superior to simpler methods, including a fully non-stationary isotropic model, which demonstrates the value of using the cylindrical distance approach to model anisotropy. The hierarchical Bayesian approach yields posterior distributions for each of the parameter fields which allows us to assess the uncertainty in their estimation. This additional source of uncertainty is then represented in the posterior distribution of ocean heat content and its trend. Our model's non-stationary representation of the mean and trend fields allows for estimating the spatial distribution of certainty regarding local warming or cooling trends. The models have been fit in a computationally efficient way using the Vecchia process, which we have shown yields a sufficiently accurate approximation to the full Gaussian process. The code in this paper is available as the R package \pkg{BayesianOHC} which is publicly available at \url{https://github.com/samjbaugh/BayesianHeatContentCode}. While this package has many utilities that are specifically useful for modeling Argo data, many of the model-fitting functions are general and can be used for fitting non-stationary Bayesian models for a variety of applications.
	
	The results presented in this paper are limited in the amount of data considered, specifically to January over the years 2007-2016 and for floats whose profiles extend to $1{,}900$m of depth. In future work, we are interested in extending the model fit to shallower floats as well as to additional months. As the heat content field is defined here as the integral to $2{,}000$m of depth, one way to incorporate floats with a shallower max depth which are nonetheless located in areas of the ocean that are deeper than $2{,}000$m would be to train an extrapolation model on the full-depth floats and then apply this model to complete the temperature profiles of the shallower-depth floats. The uncertainty in ocean heat content values induced by this extrapolation could then be incorporated into the Bayesian model structure as an additional nugget effect for the applicable observations.
	
	In this work, restricting the data to January has precluded the need to model temporal correlation and seasonal variation, while also providing the major computational advantage in that each year's likelihood function could be evaluated independently. In extending the model to all months of the year both of these challenges will need to be addressed. To model seasonal variation in the mean field, a temporal dimension could be added to the knot structure of the current spatially varying mean field. Then the number of temporal knots would need to be chosen to be large enough to ensure that the resulting spatio-temporal mean field adequately models seasonality. Our use of the Vecchia process largely remedies the computational burden of considering additional months, as for a fixed $m$ the computation time will grow linearly with the number of observations.  However, the choice of conditioning sets for the Vecchia process would need to be extended in time, and the accuracy of this extension would need to be evaluated in the spatio-temporal context. This will be investigated in future work.

	\linespread{.5}\selectfont
	
	\bibliography{AOAS1605_bib}{}

\end{document}


\section{Region Maps}

\begin{figure}[H]
				\centering
				\includegraphics[width=\linewidth]{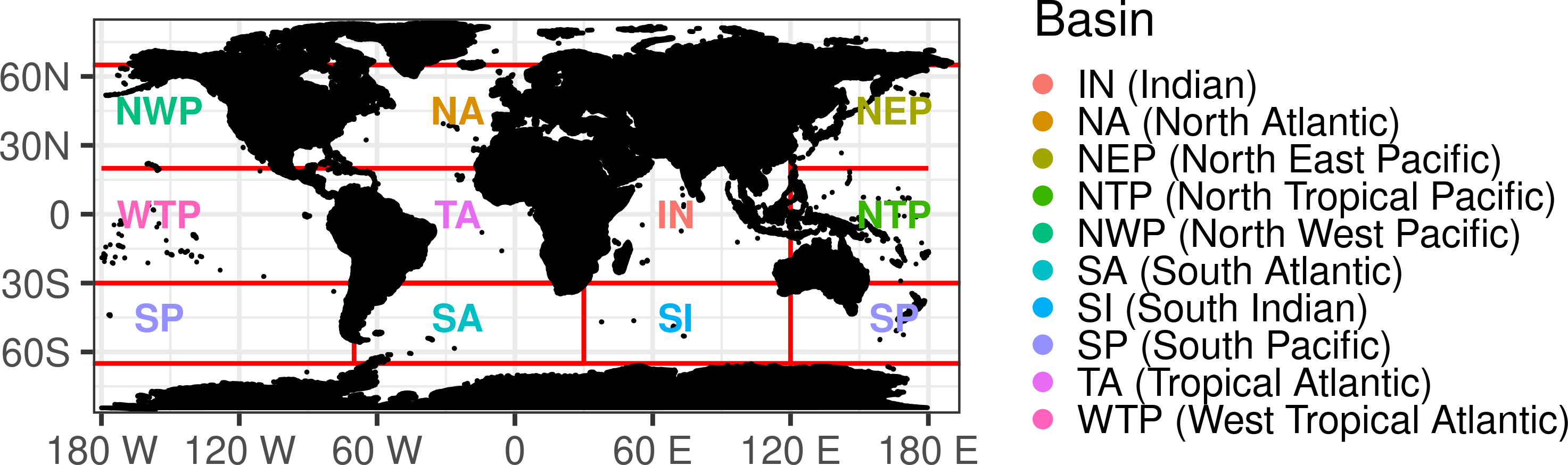}
				
				\caption{Basin definitions used for evaluation of the accuracy of the Vecchia approximation.} \label{basin_definitions}
			\end{figure}

			Figure \ref{basin_definitions} shows the latitudinal and longitudinal lines demarcating the regions used in the Vecchia approximation accuracy study in Table 1 of the main text. 
			
			\section{Comparison with Results Obtained from the Method Described in Levitus et al.}

			The approach to interpolating ocean heat content and calculating standard errors used in \cite{levitus2012world} is commonly cited in the climate community. In this method, the predicted ocean heat content value at an unobserved point is the weighted sum of observations within a fixed radius around the point. The weights are determined by an exponential function of the negative squared distance between the locations. Standard errors are calculated by computing the weighted standard deviation of observations within the radius with the assumption that the errors at different locations are independent.  
			
\begin{figure}[H]
				\centering
				\includegraphics[width=\linewidth]{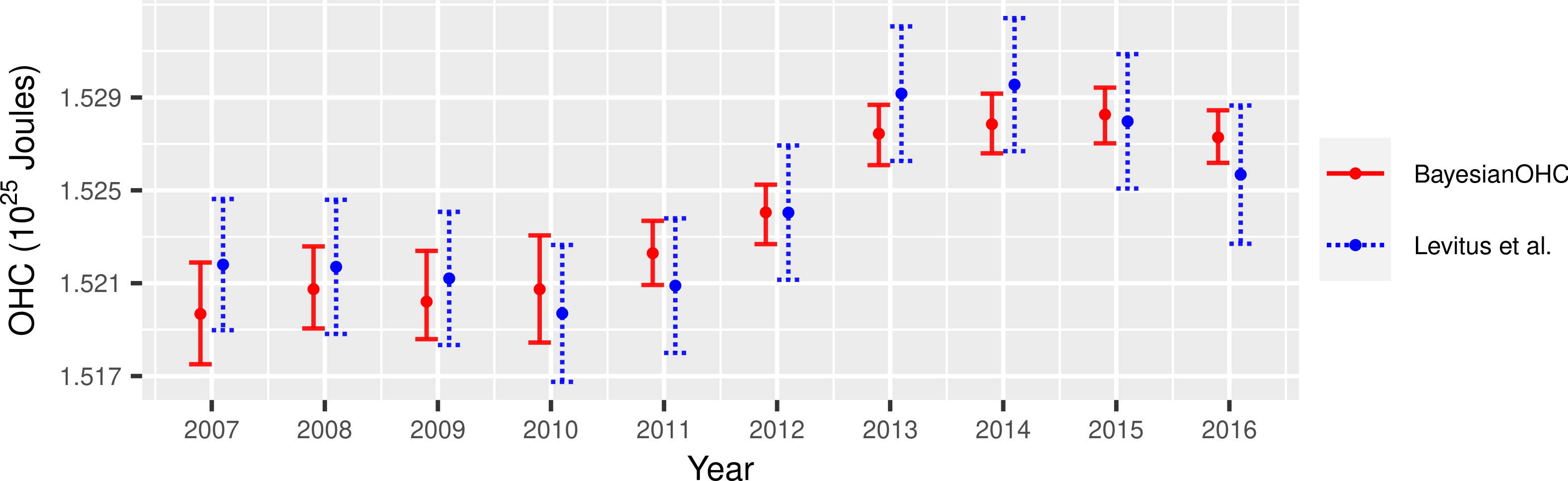}
				\caption{Comparison of predicted ocean heat content values and uncertainty intervals between the non-stationary cylindrical method and the Levitus et al. method.}
				\label{levitus_comparison}
			\end{figure}

			While Table 3 in the main text and Table \ref{lowo_cv} in the supplement show that our non-stationary Gaussian process model out-performs the Levitus et al. method at cross-validation, we are interested in how the predicted ocean heat content values and uncertainties compare between the two methods. Figure \ref{levitus_comparison} displays the predicted global OHC values and $95\%$ uncertainty intervals obtained using the \cite{levitus2012world} method along with the posterior medians and credible intervals obtained using the \pkg{BayesianOHC} fully non-stationary cylindrical distance model. After using the Levitus approach to compute global heat content values for each year, the implied trend was estimated using ordinary least-squares and the corresponding confidence interval was calculated. The resulting estimated trends and uncertainty intervals are displayed in Table \ref{trend_intervals}. 
			
			We can see that both methods are fairly similar in terms of predicting yearly heat content values, however, the standard errors using the Levitus approach seen in Figure \ref{levitus_comparison} are substantially larger than those obtained from \pkg{BayesianOHC}. This is to be expected as unlike in a Gaussian process framework, the \cite{levitus2012world} approach does not take into consideration correlations between locations and rather assumes that the errors in distinct gird-boxes are independent. On the other hand, it can be seen in Table \ref{trend_intervals} that the \pkg{BayesianOHC} credible interval is wider than the ordinary least-squares confidence interval. This is also not surprising due to fact that the uncertainty from fitting ordinary least-squares to the Levitus predictions does not take into account the uncertainty induced from the spatial interpolation, and as a result the confidence interval underestimates the uncertainty in the trend. This is in contrast to the \pkg{BayesianOHC} credible intervals which are estimated within the context of a hierarchical model, and as such take into account the uncertainty induced by the interpolation of the heat content field as well as the additional uncertainty from estimating the other parameters in the model. 
			
			Despite the wider uncertainty range in the \pkg{BayesianOHC} estimated trend, the larger median trend estimate yields a smaller p-value than the \cite{levitus2012world} least-squares fit.  If hypothesis testing was used, the \cite{levitus2012world} approach would fail to reject a null hypothesis of no trend at a one percent significance threshold whereas the \pkg{BayesianOHC} method would conclude that there is sufficient evidence that the oceans have gained heat between 2007 and 2016. This demonstrates the value of using statistical modeling to quantify ocean heat content as our model more accurately quantifies heat content and its variability with a meaningful impact on the significance of the trend.
			
			\begin{table}[ht]
				\centering
				\begin{tabular}{rlllll}
					\hline
					& 2.5\% & Median & 97.5\% & Width & p-value \\ 
					\hline
					Levitus et al. & 2.753$\times 10^{21}$ & 9.215$\times 10^{21}$ & 15.677$\times 10^{21}$ & 12.924$\times 10^{21}$ & 0.01105 \\ 
					BayesianOHC & 3.537$\times 10^{21}$ & 10.321$\times 10^{21}$ & 17.139$\times 10^{21}$ & 13.601$\times 10^{21}$ & 0.00165 \\ 
					\hline
				\end{tabular}
				\caption{Estimated trend and associated $95\%$ confidence interval for the ordinary least-squares trend fit to the heat content predictions computed using the \cite{levitus2012world} method and the median posterior trend and associated $95\%$ credible interval using the fully non-stationary \pkg{BayesianOHC} fit. All units are in J/year.}
				\label{trend_intervals}
			\end{table}

			\section{Cylindrical Kernel Convolutions}
			In this section we will show how the cylindrical kernel convolutions can be computed exactly using Gaussian error functions in the spherical dimension. We then demonstrate with a simulation study that a Gaussian function approximation to the exact convolutions is accurate over the range of longitudinal correlations present in the Argo data.
			
			\subsection{Formulas for Exact Convolutions}
			
			We can write the kernel convolution covariance described in Equation 3 in Section 3.2 of the main text as 
			\[k(\textbf{x},\textbf{y};\boldsymbol{\rho})=\phi(\textbf{x})\phi(\textbf{y}) \int_{\textbf{u}\in\mathbb{CL}}\exp\left(-d_{\textnormal{cyl}}(\textbf{x},\textbf{u};\theta_{\textnormal{lat}},\theta_{\textnormal{lon}})^2-d_{\textnormal{cyl}}(\textbf{y},\textbf{u};\theta_{\textnormal{lat}},\theta_{\textnormal{lon}})^2\right)\,d\textbf{u}\] \[=\int_{\mathbb{R}}\exp\left(\frac{d_{\textnormal{euc}}(x_{\textnormal{lat}},u_{\textnormal{lat}})^2}{\theta_{\textnormal{lat}}(x_{\textnormal{lat}})}-\frac{d_{\textnormal{euc}}(y_{\textnormal{lat}},u_{\textnormal{lat}})^2}{\theta_{\textnormal{lat}}(y_{\textnormal{lat}})}\right)\,du\int_{\mathbb{S}^1}\exp\left(\frac{d_{\textnormal{gc}}(x_{\textnormal{lon}},u_{\textnormal{lon}})^2}{\theta_{\textnormal{lon}}(x_{\textnormal{lon}})}-\frac{d_{\textnormal{gc}}(y_{\textnormal{lon}},u_{\textnormal{lon}})^2}{\theta_{\textnormal{lon}}(y_{\textnormal{lon}})}\right)\,du\]
			
			Let $G^{b}_a(x,y;\theta_x,\theta_y)=\int_a^b\exp\left(-\frac{(u-x)^2}{\theta_x}-\frac{(u-y)^2}{\theta_y}\right)\,du$ denote the Gaussian convolution with generic parameters and limits of integration. Through completing the square we can see that	\[G_a^b(x,y;\theta_x,\theta_y)=\exp\left(\frac{(y-x)^2}{\theta_x+\theta_y}\right)\int_a^b\frac{u-\frac{x\theta_x+y\theta_y}{\theta_x+\theta_y}}{\theta_x\theta_y/(\theta_x+\theta_y)}\,du\] 
			\begin{equation}
				=\sqrt{\pi}\sqrt{\theta_{xy}}\exp\left(\frac{(y-x)^2}{\theta_x+\theta_y}\right)\left[\Phi\left(b;\frac{x\theta_x+y\theta_y}{\theta_x+\theta_y},\sqrt{\theta_{xy}}\right)-\Phi\left(a;\frac{x\theta_x+y\theta_y}{\theta_x+\theta_y},\sqrt{\theta_{xy}}\right)\right] \label{regionint}
			\end{equation}
			
			\noindent where $\theta_{xy}=\frac{\theta_x\theta_y}{\theta_x+\theta_y}$ and $\Phi$ denotes the normal cumulative density function (CDF) with mean and variance given as the first and second arguments respectively. By taking the limits of Equation \ref{regionint} we can see that the latitudinal dimension term of the kernel convolution has a simple closed-form expression: \[\int_{\mathbb{R}}\exp\left(\frac{d_{\textnormal{euc}}(x_{\textnormal{lat}},u_{\textnormal{lat}})^2}{\theta_{\textnormal{lat}}(x_{\textnormal{lat}})}-\frac{d_{\textnormal{euc}}(y_{\textnormal{lat}},u_{\textnormal{lat}})^2}{\theta_{\textnormal{lat}}(y_{\textnormal{lat}})}\right)\,du=G^{\infty}_{-\infty}(x_\textnormal{lat},y_\textnormal{lat};\theta_\textnormal{lat}(x_\textnormal{lat}),\theta_\textnormal{lat}(y_\textnormal{lat}))\] \begin{equation}
				=\sqrt{\pi}\sqrt{\frac{\theta_\textnormal{lat}(x_\textnormal{lat})\theta_\textnormal{lat}(y_\textnormal{lat})}{\theta_\textnormal{lat}(x_\textnormal{lat})+\theta_\textnormal{lat}(y_\textnormal{lat})}}\exp\left(-\frac{1}{2}\frac{(x_{\textnormal{lat}}-y_{\textnormal{lat}})^2}{\theta_\textnormal{lat}(x_\textnormal{lat})+\theta_\textnormal{lat}(y_\textnormal{lat})}\right) \label{euc_gaussian}
			\end{equation}
			
			For the cylindrical integral we first rotate the locations $x$ and $y$ to $x'=0$ and $y'=d_{\textnormal{gc}}(x,y)$ which preserves the great-circle distance between the two points. Since $u\in [-\pi,\pi]$ the first distance term becomes $d_{\textnormal{gc}}(x',u)=(x'-u)^2$. For the second distance term, we can split the integral between $u\in (y-\pi,\pi)$ where $d_{\textnormal{gc}}(y',u)=(y'-u)^2$ and $u\in (-\pi,y-\pi)$ where $d_{\textnormal{gc}}(x',u)=(y'-2\pi-u)^2$. This yields \[\int_{-\pi}^\pi \exp\left(-\frac{d_{\textnormal{gc}}(x',u)}{\theta_x}-\frac{d_{\textnormal{gc}}(y',u)}{\theta_y}\right)\,du\] \[=\int_{-\pi}^{y'-\pi} \exp\left(-\frac{(x'-u)^2}{\theta_x}-\frac{(y'-2\pi-u)^2}{\theta_y}\right)\,du+\int_{y'-\pi}^\pi \exp\left(-\frac{(x'-u)^2}{\theta_x}-\frac{(y'-u)^2}{\theta_y}\right)\,du\] \begin{equation}
				=G_{-\pi}^{u'-\pi}(x',y'-2\pi;\theta_x,\theta_y)+G_{y'-\pi}^\pi(x',y';\theta_x,\theta_y). \label{cyl_exact}
			\end{equation} Then by substituting in $x_{\textnormal{lon}}$ and $\theta_{\textnormal{lon}}(x_{\textnormal{lon}})$ into the above equation the spherical convolution can be computed using two evaluations of Equation \ref{regionint}. This is a closed form expression involving the use of the Gaussian CDF, which has pre-computed evaluation functions in most programming languages. The $G$ function described above is available in the \pkg{BayesianOHC} package as the function 	``gaussian\_convolution'' and the exact convolutions in the longitudinal and latitudinal dimensions are available as ``cylindrical\_correlation\_exact'' and ``euclidean\_correlation\_convolution'' respectively. We note that in the implementations we follow \cite{Paciorek2006} and scale the convolutions by $\frac{\sqrt{2}}{\theta_{\textnormal{lon}}(x_{\textnormal{lon}})^{1/4}\theta_{\textnormal{lon}}(x_{\textnormal{lon}})^{1/4}\sqrt{\pi}}$ and $\frac{\sqrt{2}}{\theta_{\textnormal{lat}}(x_{\textnormal{lat}})^{1/4}\theta_{\textnormal{lat}}(x_{\textnormal{lat}})^{1/4}\sqrt{\pi}}$ respectively to ensure that the correlation function between a location and itself evaluates to one. 
			
			\subsection{Simulation Study Justifying the Use of Gaussian Approximation}
			
			While Gaussian CDFs are faster to compute than estimating the integrals in the convolution numerically, evaluating the pre-computed CDF still takes longer to compute than the latitudinal convolutions using Equation \ref{euc_gaussian}. One would expect that when the correlation length scales are small relative to the circumference of the Earth, the circular convolutions will behave similarly to the Euclidean convolutions and the following alteration of Equation \ref{euc_gaussian} could be used to approximate the convolution: \begin{equation}
				=\sqrt{\pi}\sqrt{\frac{\theta_\textnormal{lon}(x_\textnormal{lon})\theta_\textnormal{lon}(y_\textnormal{lon})}{\theta_\textnormal{lon}(x_\textnormal{lon})+\theta_\textnormal{lon}(y_\textnormal{lon})}}\exp\left(-\frac{1}{2}\frac{d_{\textnormal{gc}}(x_{\textnormal{lon}},y_{\textnormal{lon}})^2}{\theta_\textnormal{lon}(x_\textnormal{lon})+\theta_\textnormal{lat}(y_\textnormal{lon})}\right). \label{cyl_gaussian}
			\end{equation} 
			
\begin{figure}[H]
				\centering
				\includegraphics[width=\linewidth]{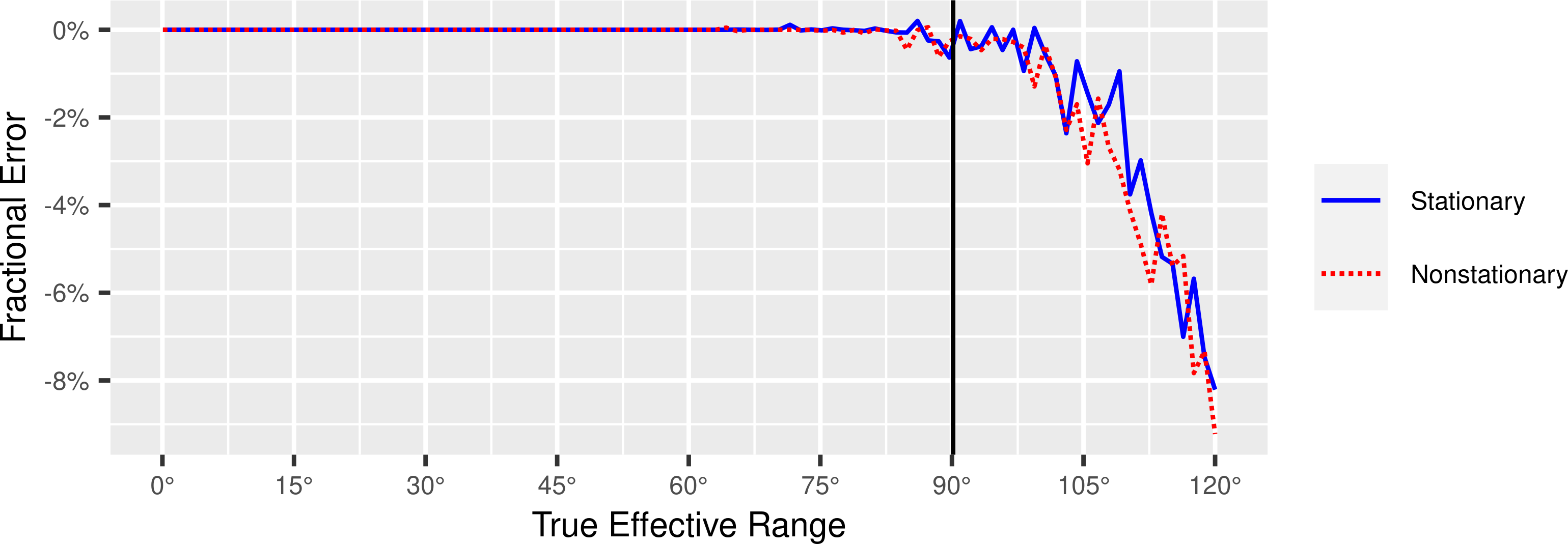}
				\caption{Fractional errors of the exact method over the Gaussian approximation. The horizontal line is at the $99^{\textnormal{th}}$ percentile of the initial condition distribution of $\theta_{\textnormal{lon}}$. }
				\label{cylind_sim}
			\end{figure}

			In order to determine the size of the correlation lengths that would be required for Equation \ref{cyl_gaussian} to be a sufficient approximation for the exact convolutions, we performed a simulation to compare the two correlation functions on data randomly generated according to known parameters. First, one hundred locations $x$ were simulated from the uniform distribution of the unit circle. Then a dense grid of $\theta$ was generated corresponding to an effective range length between zero and $120^\circ$ degrees, where ``effective range'' here is the distance at which the correlation becomes less than $0.05$. For each value of $\theta$ two types of correlation structures were simulated: a stationary structure where $\theta(x)=\theta$ for all locations $x$ and a non-stationary structure $\theta(x)=\cos(x)\theta/2+\theta$. In the non-stationary structure, local range parameters vary between $\theta/2$ and $3\theta/2$ with $\theta$ being the mid-point of the parameter field. For each value of $\theta$ one hundred sets of observations were randomly generated using covariance matrices generated from the exact convolution method for both the stationary and non-stationary correlation structures. Then for each simulated set of data, maximum likelihood estimates (MLEs) were computed according to both the exact method and the approximation method, and the average was computed over the 100 simulations. Finally fractional errors of the approximation method MLEs over the exact method MLEs where calculated, specifically $\frac{\hat{\theta}_{\textnormal{approx}}-\hat{\theta}_{\textnormal{exact}}}{\hat{\theta}_{\textnormal{exact}}}$ where $\hat{\theta}$ refers to the MLEs averaged over the one hundred simulations. Figure \ref{cylind_sim} shows the fractional errors for the stationary and non-stationary correlation structures over each value of $\theta$ displayed in terms of the effective range in degrees.

			As we can see, the approximation is nearly exact for effective range parameters less than $80^\circ$, beyond which the approximation method begins to yield increasingly larger negative errors. This trend is the same between the stationary and non-stationary simulations. The fact that the approximation does increasingly worse at high values of the range parameter is to be expected, as at this point the value of the exact convolution will begin to diverge from the Gaussian approximation. 
			
			The horizontal line in Figure \ref{cylind_sim} shows the $99^{\textnormal{th}}$ percentile of $\theta_{\textnormal{lon}}$ in the fitted fully non-stationary cylindrical distance model described in the main text, which corresponds to an effective range of $90.13$. At this value the stationary and non-stationary fractional errors are $-.6\%$ and $-.2\%$ respectively. The small error of the approximation at the largest correlation length scales present in the ocean heat content field demonstrates that the Gaussian approximation to the exact spherical convolutions is sufficient for modeling ocean heat content. We note that at the initial configuration used for the MCMC configuration, the $99^{\textnormal{th}}$ quantile of the effective ranges was $65.40^\circ$ at which point the fractional errors from the simulation are less than a hundredth of a percent. Due to its sufficiency and computational advantages, the approximation was used in generating the results displayed in the main text, and is available in the \pkg{BayesianOHC} package as the function ``cylindrical\_correlation\_gaussian''.

			\section{Windowed Cross-Validation Results}
			
\begin{figure}[H]
				\centering
				\includegraphics[width=.7\linewidth]{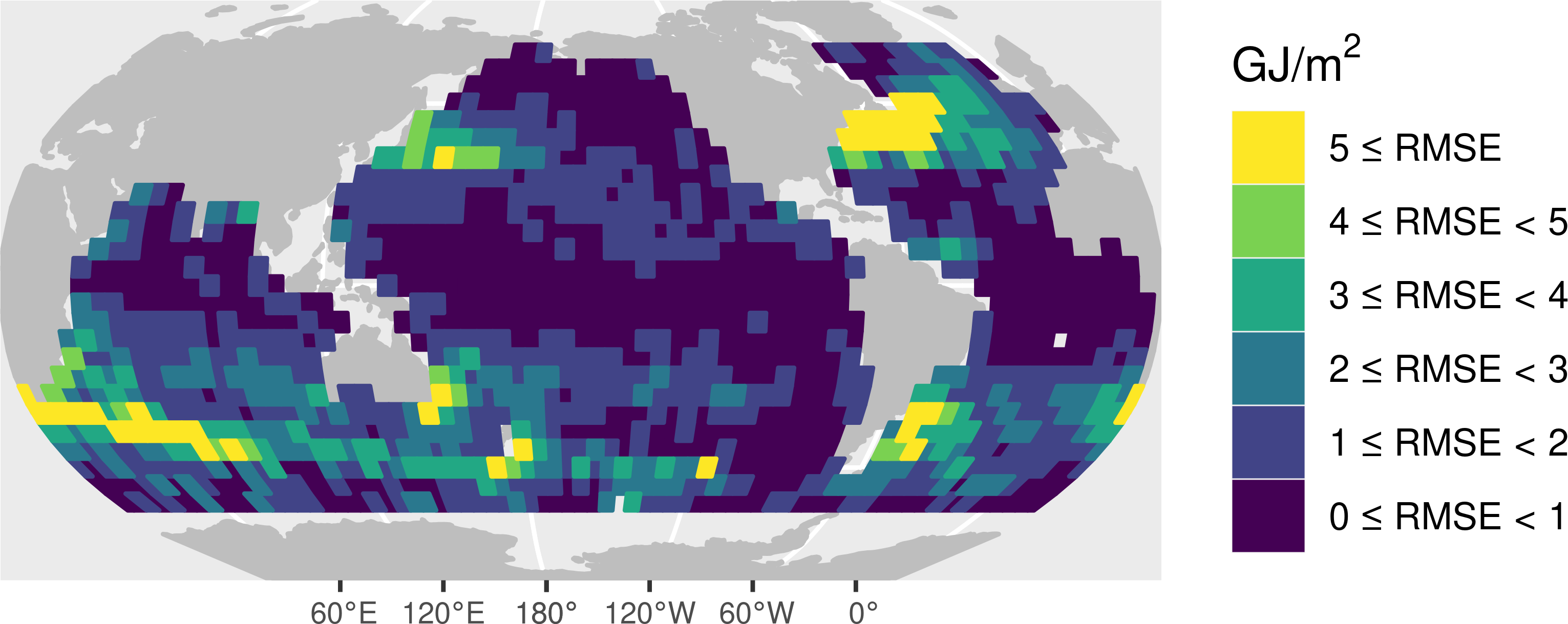}
				
				\caption{Average windowed cross validation MAE values over all years for each grid-cell of a $5^\circ\times 5^\circ$ grid.} \label{lofo_grid}
			\end{figure}

			In addition to the leave-one-float-out (LOFO) cross-validation study in Table 8 of the main text, we performed cross-validation on the same set of models using a windowed approach, where for each observation the observed locations within a $2^\circ\times 2^\circ$ window around the location are withheld in order to get a sense for how well the models predict at unobserved locations. The mean absolute error (MAE) scores averaged over each $5^\circ\times 5^\circ$ grid-box are shown in Figure \ref{lofo_grid}. Similar patterns are seen in the analogous Figure 7 for LOFO cross-validation although the LOFO MAE scores are $10.9\%$ larger on average. 
			
			The windowed cross-validation mean absolute error (MAE), root-mean-squared error (RMSE), and continuous ranked probability scores (CRPS) are displayed in Table \ref{lowo_cv} along with percent improvements over the \citet{levitus2012world} reference model. Similar trends can be seen as in Table 3 in the main text, with the reference model generally under-performing the other models and the fully non-stationary anisotropic model generally out-performing the other models.
			
			\begin{table}[ht]
				\centering
				\begin{tabular}{rllll}
					\hline
					& Package & MAE & RMSE & CRPS \\ 
					\hline
					Fully Non-stationary & BayesianOHC & 1.451 (11.11\%) & 2.375 (12.09\%) & 1.174 (12.13\%) \\ 
					Stationary $\sigma^2/\phi$ & BayesianOHC & 1.458 (10.63\%) & 2.357 (12.76\%) & 1.175 (12.03\%) \\ 
					Stationary $\theta_{\textnormal{lat}}$ & BayesianOHC & 1.471 (9.83\%) & 2.403 (11.05\%) & 1.188 (11.03\%) \\ 
					Nonstationary Isotropic & BayesianOHC & 1.475 (9.62\%) & 2.404 (11.02\%) & 1.195 (10.54\%) \\ 
					Stationary $\theta_{\textnormal{lon}}$ & BayesianOHC & 1.499 (8.13\%) & 2.427 (10.18\%) & 1.212 (9.29\%) \\ 
					Stationary Spatiotemporal & GpGp & 1.534 (6.01\%) & 2.469 (8.60\%) & 1.432 (-7.19\%) \\ 
					Stationary $\phi$ & BayesianOHC & 1.541 (5.56\%) & 2.446 (9.48\%) & 1.246 (6.72\%) \\ 
					Stationary Isotropic & GpGp & 1.544 (5.41\%) & 2.477 (8.31\%) & 1.445 (-8.15\%) \\ 
					Fully Stationary & BayesianOHC & 1.610 (1.35\%) & 2.479 (8.26\%) & 1.298 (2.80\%) \\ 
					Levitus et al. & BayesianOHC & 1.632 (0.00\%) & 2.702 (0.00\%) & 1.336 (0.00\%) \\ 
					\hline
				\end{tabular}
				\caption{Windowed cross-validation scores for the fully non-stationary cylindrical distance model, restrictions to stationarity in each of the parameter fields, models using the chord-length distance, and a stationary spatio-temporal model. Chordal distance models and the spatio-temporal model are isotropic. }
				\label{lowo_cv}
			\end{table}
			
			\bibliography{AOAS1605_bib}{}